\documentclass[12pt,preprint]{aastex}

\shortauthors{Li \& Yee}
\shorttitle{The pFoF Algorithm}

\begin{document}

\title{Finding Galaxy Groups In Photometric Redshift Space: the Probability Friends-of-Friends (pFoF) Algorithm}

\author{I.H. Li}
\email{tornado@astro.utoronto.ca}
\author{H.K.C. Yee}
\email{hyee@astro.utoronto.ca}
\affil{50 St. George Street, Department of Astronomy \& Astrophysics, University of Toronto,
    Toronto, ON, Canada, M5S 3H4 }

\begin{abstract}
   We present a structure finding algorithm designed to identify galaxy groups in photometric redshift data sets: the probability friends-of-friends (pFoF) algorithm. This algorithm is derived by combining the friends-of-friends algorithm in the transverse direction and the photometric redshift probability densities in the radial dimension. The innovative characteristic of our group-finding algorithm is the improvement of redshift estimation via the constraints given by the transversely connected galaxies in a group, based on the assumption that all galaxies in a group have the same redshift. Tests using the Virgo Consortium Millennium Simulation mock catalogs allow us to show that the recovery rate of the pFoF algorithm is larger than 80\% for mock groups of at least $2\times10^{13}M_{\sun}$, while the false detection rate is about 10\% for pFoF groups containing at least $\sim8$ net members. Applying the algorithm to the CNOC2 group catalogs gives results which are consistent with the mock catalog tests. From all these results, we conclude that our group-finding algorithm offers an effective yet simple way to identify galaxy groups in photometric redshift catalogs. 
 \end{abstract}


\section{Introduction}
  Galaxy groups are sites where local galaxy number density is relatively higher than the field. 
The majority ($\sim60\%$) of galaxies lies in groups \citep[e.g.,][]{2004MNRAS.348..866E,2006ApJS..167....1B,2006AN....327..365T}, so that galaxy groups provide an excellent location to study the effect of local environment on galaxy formation and evolution. 
Unlike galaxy clusters, galaxy groups, especially those at high redshift, are not easy to detect because of their smaller size and the significantly lower hot gas density. 
The current published galaxy group catalogs are constructed based on large-scale galaxy redshift surveys using automated group finding schemes. 
The techniques include the popular friends-of-friends algorithm \citep[e.g.,][]{1983ApJS...52...61G, 2005ApJ...630..759M, 2004MNRAS.348..866E} and the Voronoi partition technique \citep[e.g.,][]{2005ApJ...625....6G}. 
Most of these catalogs list galaxy groups either in the nearby Universe ($ z < 0.1$) or over a small sky area. 
Galaxy groups of large sample sizes in intermediate and higher redshift space still remain largely unexplored.

  Up to now, most structure finding techniques use spectroscopic redshift or simulated catalogs, both containing accurate three-dimensional position information. 
With the development of the photometric redshift method, the approximate redshifts of all galaxies in a photometric multi-band survey can be obtained without the time-consuming spectroscopic measurements. 
The photometric redshift method involves either SED (spectral energy distribution) fitting \citep[e.g.,][]{2000A&A...363..476B,2003ApJ...586..745C,2004ApJ...600L.167M, 2004ApJS..155..243W,2005ApJ...619L..31B,2006ApJS..162...20B} or the use of a spectral `training set' to compute the photometric redshift via an empirical polynomial of galaxy colors and magnitudes \citep[e.g.,][]{1995AJ....110.2655C, 2005ApJS..158..161H}. 
Since the redshifts are derived from broadband galaxy colors rather than from spectra, the photometric redshift method can estimate the redshift of objects which are too faint for spectroscopy. 
On the other hand, photometric redshifts have larger uncertainties by a factor of $50-100$ than those measured from spectroscopy. 
Due to the less accurate distance information in photometric redshift catalogs, the main problem of structure finding is the blurring of configurations in redshift space, producing unrealistic or elongated structures caused by the large photometric redshift uncertainties \citep{2004MNRAS.349..425B}. 
Even with excellent photometric redshift estimation ($\sigma_z \sim 0.03$), the structures on the small scale will still be largely smeared out. 
Furthermore, projection effects make the subtraction of foreground and background galaxy contamination important in analyzing structures found using photometric redshift. 

  In order to overcome some of these problems, we propose here a method of finding galaxy groups in photometric redshift catalogs. 
The knowledge of galaxy photometric redshift uncertainty or probability density is required for this method. 
This group finding methodology is based on the idea of the standard friends-of-friends algorithm in the transverse direction, but takes into account the photometric redshift probability density to determine the friendship in the radial direction. 
We describe the photometric redshift technique and the error estimation for individual galaxies in $\S$\ref{sec_photoz}, and present our photometric sample selection criteria in $\S$\ref{sec_sample}. 
The group-finding parameters and the algorithm are detailed in $\S$\ref{sec_pFoFpar} and $\S$\ref{sec_pFoF}. 
The basic properties of galaxy groups are quantified in $\S$\ref{sec_grp}. 
This algorithm is tested in $\S$\ref{sec_test} using mock catalogs constructed from the \it Virgo Consortium Millennium Simulation \rm \citep{2005Natur.435..629S}, and applied to the real observed groups in the \it Canadian Network for Observational Cosmology Survey \rm \citep[CNOC2;][]{2000ApJS..129..475Y} in $\S$\ref{cnoc2grp}.
 Finally, we present a summary in $\S$\ref{sec_conclusion}. 
The analyses of galaxy group samples from a number of surveys will be presented in future papers. 
We adopt the standard cosmological parameters of $H_0$=70 km/s/Mpc, $\Omega_m=0.3$, and $\Omega_{\Lambda}=0.7$.

\section{Empirical Photometric Redshift \label{sec_photoz}}
\subsection{Training set}
We estimate photometric redshift using the empirical fitting technique \citep[e.g.,][]{1995AJ....110.2655C, 2005ApJS..158..161H}. 
We express the galaxy redshift as a quadratic polynomial in magnitudes and colors:
	\begin{equation} \label{eq:photoz}
        z_i = a_0 + \sum a_i m_i + \sum a_{ij} (m_i- m_{j})^2, 
	\end{equation}
where $m_i$ and $m_{j}$ are the passband magnitudes and $a_0, a_i$ and $a_{ij}$ are the constant term and the coefficients associated with the magnitudes and colors, respectively. 
The coefficients in equation (1) can be derived by fitting a training set, a catalog which contains both galaxy redshift and photometry information. 

  Our training set is constructed using data from the \it Red-Sequence Cluster Survey \rm \citep[RCS;][]{2005ApJS..157....1G} in four CNOC2 survey patches \citep{2000ApJS..129..475Y} and the GOODS/HDF-N field \citep{2004ApJ...600L..93G}. 

   The RCS was designed to find galaxy clusters at $0.4 < z < 1.4$ using the cluster red-sequence method with $R_c$ and $z'$ filters. 
It includes 22 widely separated patches covering a total area of 90 deg$^2$, observed with the CFHT 12K camera in the Northern Hemisphere and the CTIO 4m MOSAIC II camera for the Southern Sky. 
The RCS follow-up covers 33.6 deg$^2$ (corresponding to about 75 \%  of the CFHT RCS fields) observed with the 12k camera in $B$ and $V$. 
The photometry has been  carried out  using \it PPP \rm \citep{1991PASP..103..396Y, 1996ApJS..102..269Y} and internally calibrated using star colors and galaxy counts. 
It has also been cross-checked with star colors and counts from \it Sloan Digital Sky Survey (SDSS) \rm Data Release 3 \citep[DR3;][]{2005AJ....129.1755A}. 
The RCS follow-up sample is 100\% complete to $R_c$ $\sim 24.2$. 
Further details on the data and on the photometric reduction can be found  in \cite{2005ApJS..158..161H}. 
The CNOC2 survey  covers over 1.5 deg$^2$ of sky with a total sample of $\sim6200$ galaxies (up to $z \sim 0.55$) with $R_c \leq 22.0$r; 1727 of these galaxies overlap with the RCS sample. 

  The GOODS HDF-N field allowed us to extend our training set sample to larger redshifts. 
The GOODS is a survey based on multi-band imaging data obtained with the Advanced Camera for Surveys (ACS) on the Hubble Space Telescope (HST). 
It covers two fields, HDF-N and CDF-S, with a total area of about 320 arcmin$^2$, a $5\sigma$ limiting magnitude in the $R$ passband (on the AB system) of 26.6, and a redshift range from 0.5 to 1.5. 
We have used publicly available $BVRz'$ photometry \citep{2004AJ....127..180C} and spectroscopic redshift \citep{2004AJ....127.3121W,2004AJ....127.3137C} for 2661 galaxies in the HDF-N field. 
To match the RCS zero point, the GOODS magnitudes have been corrected following \citet{2005ApJS..158..161H}. 
As a whole, our training set contains 3,988 galaxies observed in $BVR_cz'$ up to $z\sim 1.4$. 
The photometry uncertainties in each passband are $\Delta B\sim 0.04$, $\Delta V\sim 0.04$, $\Delta R_c\sim 0.02$, and $\Delta z'\sim 0.04$. 
Further details on the properties of this training sample can be found in  \cite{2005ApJS..158..161H}.

\subsection{Photometric Redshift Estimation and Associated Error}
\begin{figure}
        \includegraphics[width=12.7cm]{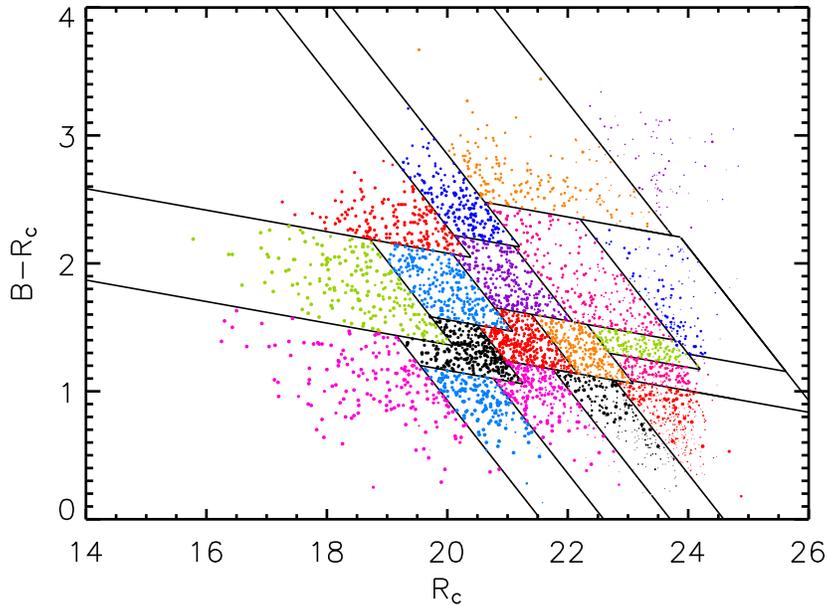}
        \caption[Color-magnitude cells in the empirical photometric fitting method]{The training set galaxies are classified into 19 color-magnitude cells in our empirical photometric fitting method. The slopes for the two sets of parallel lines are -0.084 and -0.60 to mimic the rough differentiation of different types of galaxies at various redshifts. \label{cells}}
        \end{figure}

To minimize the dispersion between photometric and spectroscopic redshifts, we separate the training set galaxies into 19 color-magnitude cells in the observed frame (Fig. \ref{cells}) to differentiate roughly different types of galaxies and different redshifts, because galaxies at high redshift tend to be fainter and redder.
To create these cells, we first sort the training set galaxies by magnitude and color, so that each cell is created starting from the region where bright and red galaxies are on the observed color-magnitude diagram. 
We use slopes of -0.084 and -0.60 for the two sets of parallel lines to create the cells. 
The slope of -0.084 is chosen based on the red sequence slope at $z\sim 0.4$ in $B-R_c$, and the other slope is determined according to the galaxy distribution for different redshift bins on the color-magnitude diagram. 
We let each cell grow by $\Delta (B-R_c)=0.1$ and $\Delta R_c=0.1$ in each step until it contains at least 160 training set galaxies. 
Galaxies are distributed into the cells according to their colors and magnitudes. 
The coefficients of Equation \ref{eq:photoz} are obtained by a linear regression method in each color-magnitude cell using the training set galaxies. 
These coefficients are then applied to those galaxies in the same color-magnitude cell to estimate their redshifts. 

  We adopt the method in \cite{2005ApJS..158..161H} to estimate photometric redshift uncertainties. 
To estimate the photometric redshift uncertainties due to fitting, we bootstrap the training set galaxies in each color-magnitude cell 300 times with the assumption of perfect photometry for each galaxy. 
On the other hand, to evaluate the contribution from photometric uncertainties, we use a Monte-Carlo method to simulate galaxy magnitudes in each passband for 300 draws with Gaussian photometry uncertainties assumed. 
With these $300\times300$ realizations, we build the photometric redshift probability density of each galaxy and take the r.m.s. dispersion as the photometric redshift uncertainty for the galaxy. 
The photometric redshift of each galaxy is the median value of these $300\times300$ executions. 

  To investigate how well the empirical photometric redshift uncertainties resemble the true ones, we define the empirical photometric redshift uncertainty $\sigma_{emp}$ as the median empirical photometric redshift uncertainties of the training set galaxies in a color-magnitude cell. 
We compute the dispersion between photometric and spectroscopic redshifts in the same cell and take it as the true uncertainty, denoted as $\sigma_{true}$. 
We find that there is a linear correlation between $\sigma_{emp}$ and $\sigma_{true}$, but not of unity slope. 
Therefore, we scale the empirical photometric redshift uncertainties and the probability densities by a factor of $\sigma_{true}/\sigma_{emp}$ in each color-magnitude cell. 

\begin{figure}
        \includegraphics[width=8.2cm]{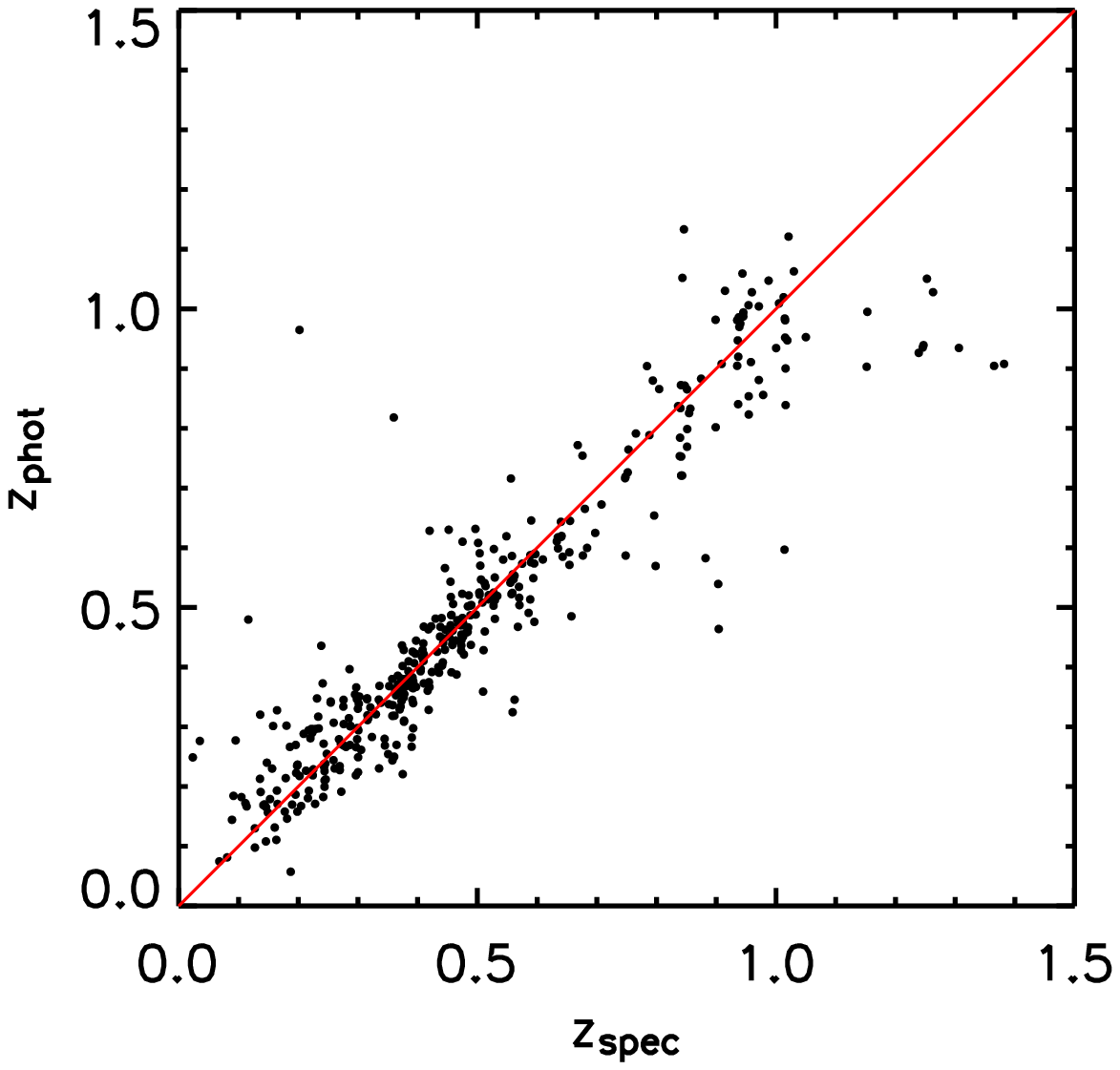}  
        \includegraphics[width=8.2cm]{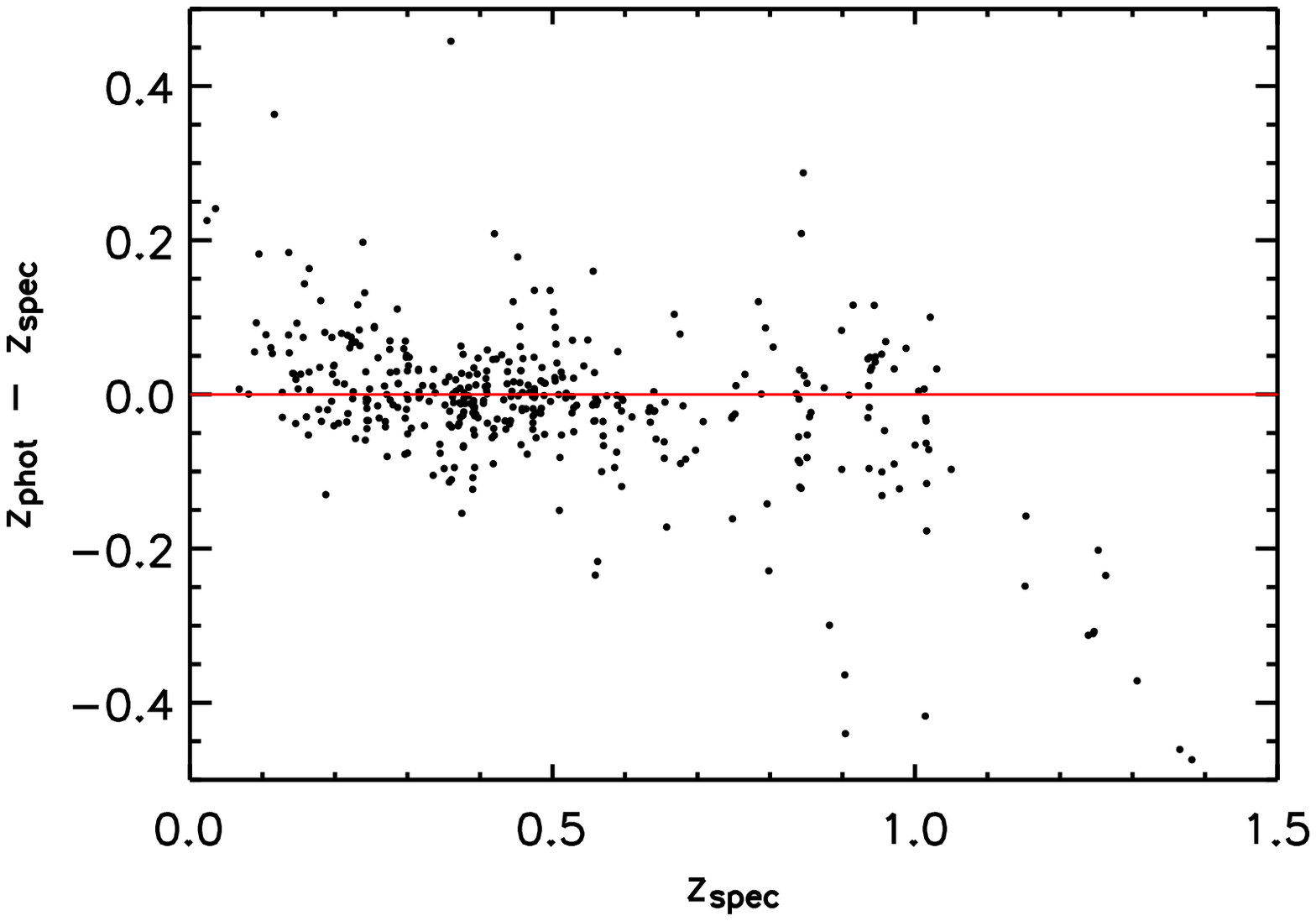}  
        \caption[Comparison between spectroscopic and photometric redshift]{The comparison between spectroscopic and photometric redshifts for 398 galaxies in $BVRz'$ trained by 3590 training set galaxies quadratically. The dispersion in redshift difference is $\sim 0.060$ at $0 \leq z < 1$. \label{subset}}
	\end{figure}

\begin{figure} 
        \includegraphics[width=8.2cm]{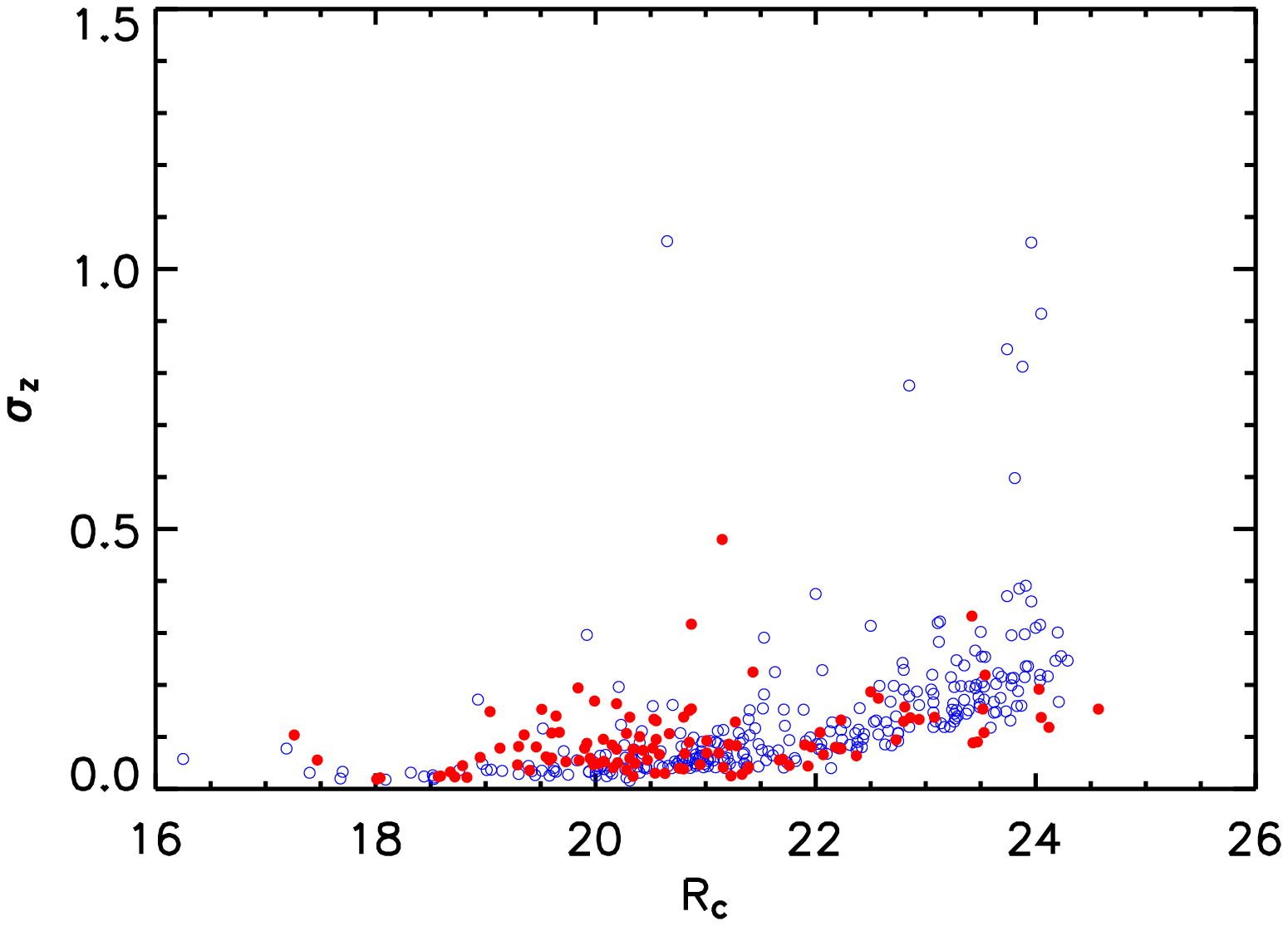}   
        \includegraphics[width=8.2cm]{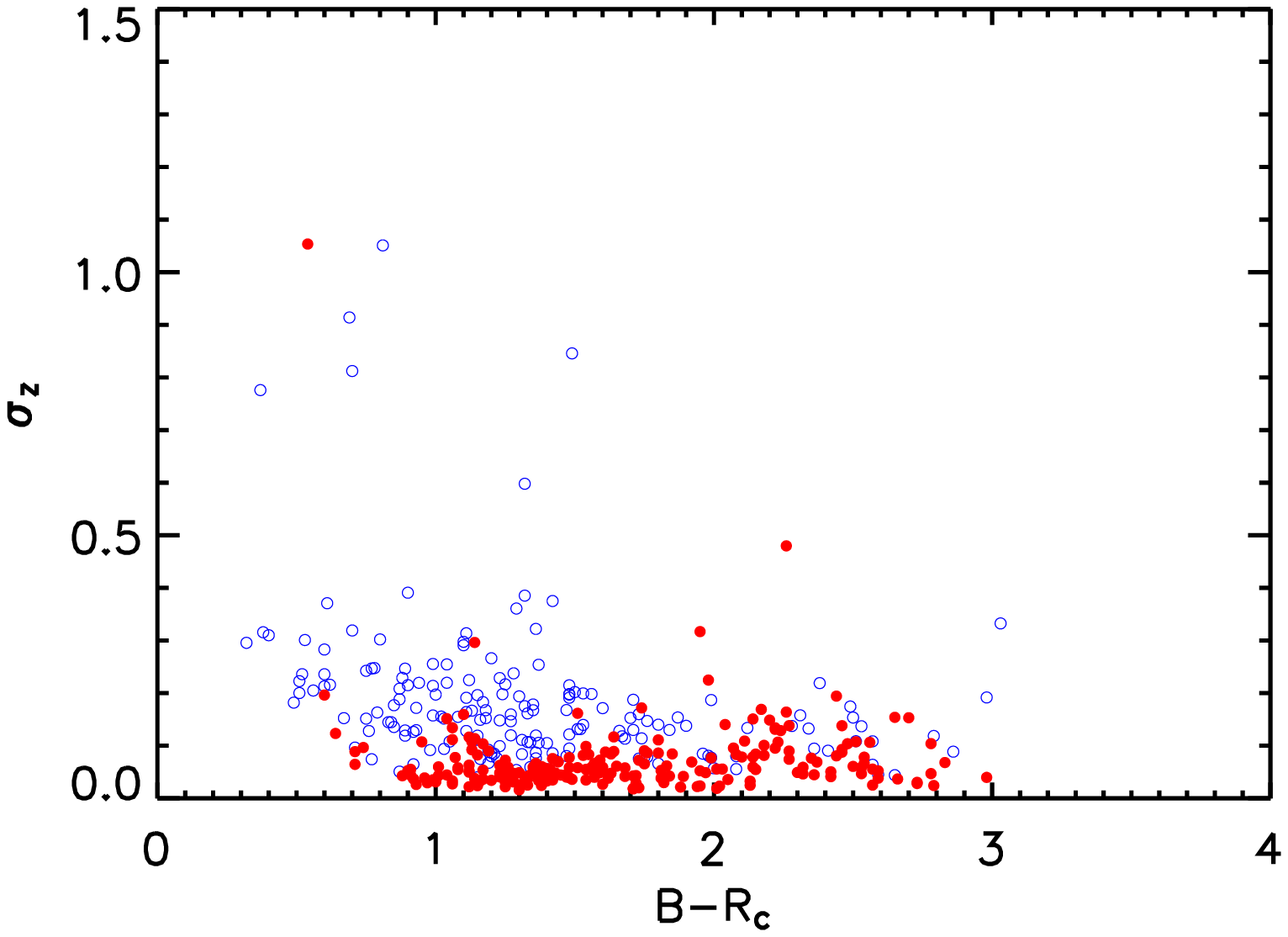} 
        \caption[Photometric redshift uncertainties as a function of magnitude and color]{\it Left\rm: The empirical photometric redshift uncertainties as a function of magnitude for the 398 control test galaxies. 
The filled circles represent red galaxies ($B-R_c \geq 1.8$) and open ones symbolize blue galaxies ($B-R_c < 1.8$).
\it Right\rm: Similar to the left but as a function of $B-R_c$ color. Filled circles are for bright galaxies ($R_c < 21.5 $) and open circles represent faint galaxies ($R_c \geq 21.5$).
        \label{errest}}
        \end{figure}

	We drop one of every ten galaxies in our training set (398 galaxies in total) and we estimate the redshift of these galaxies using the remaining training set galaxies (3590 galaxies in total), so that these two sets are independent, ensuring an unbiased estimation of the performance of our photometric redshift technique. 
The comparison of photometric redshift and spectroscopic redshift for this subset is illustrated in Fig. \ref{subset}. 
The dispersion of $\Delta z = z_{phot}-z_{spec}$ is $\sim 0.060$ for these 398 galaxies using $BVR_cz'$ photometry for $0\leq z_{spec} < 1$. 
The photometric redshift uncertainties computed using the technique described above are shown in Fig. \ref{errest} as functions of galaxy magnitude and color.
We note that the computed photometric redshift uncertainties increase for fainter and bluer galaxies.
We also apply the solutions to all the galaxies in the training set and find that redshift uncertainties increase for galaxies at higher redshift, with $\Delta z \sim 0.060$ and $\sim 0.134$, for galaxies at $0.3 \leq z < 0.6$ and $0.6 \leq z < 0.9$, respectively.  

\section{The Completeness Weight \label{sec_sample}}
  Even though the photometric redshift technique can be used to estimate a redshift for a large number of galaxies economically, the method may fail for extremely faint galaxies and galaxies with unreliable redshift. Thus, these galaxies should be excluded from the sample.
Galaxy counts must be corrected to account for such rejections. 
The selection of galaxies in a photometric redshift catalog can be based on (1) photometric redshift ranges which allow the 4000\AA~break to be within one of the pass bands, and (2) the total probability within a desired redshift range to ensure the quality of photometric redshift measurement.

  We set the redshift range to be $0.02 \leq z < 1.4$, where the upper photometric redshift limit is due to the passband wavelength coverage for the 4000\AA~break in our training set. 
We also select galaxies whose total probability within $3\sigma_{z_{cut}}$ of its central photometric redshift is greater than 99.7\%, where $\sigma_{z_{cut}}$ is set as $\sigma_{z_{cut}}=0.2(1+z)$. 

  As we select whether an object is in the sample or not, a completeness correction weight $w_i$ is assigned to each galaxy. 
Since we find that both red and blue galaxies have similar completeness correction weights, 
the completeness factor is estimated using the ratio of the total galaxy number within $\Delta m_{R_c}=0.1$ magnitude bin to the total galaxy number satisfying our selection in the same magnitude bin.
In general, this completeness weight becomes larger for fainter galaxies. 
Therefore, we set a nominal apparent magnitude cutoff based on where $w_i=2$ to avoid galaxies of high weights, if this apparent magnitude cutoff is brighter than the limiting magnitude of the sample. 

\section{Parameters for the Friendship  \label{sec_pFoFpar}}
We develop a group-finding algorithm using photometric redshift. We follow the idea of the well-known friends-of-friends algorithm in angular separation; however, we consider the conditional photometric redshift probability in the redshift direction.

\subsection{The 2D Linking Length, D0}
\indent  The standard friends-of-friends algorithm \citep[FoF;][]{1983ApJS...52...61G} identifies over-dense regions by looking for galaxies closer to one another than a given cutoff separation. 
A group forms from a seed galaxy. Galaxies satisfying the linking criterion to this seed galaxy are linked together. 
A galaxy group is defined by the chains of such finding procedures using every linked galaxy as a new seed. We adopt this linking idea in our algorithm to search for group members in the transverse direction. 
Given a fixed 2D reference linking length \it $D0_{xy}$\rm~at $z$=0, the linking length used to unite galaxies should be scaled as $D0_{xy}/(1+z)$ for the sake of forming groups of similar over-density. 
However, in an apparent-magnitude limited survey, criteria based on the distance between galaxies have to consider the variation of the mean galaxy separation with redshift \citep{2002ApJ...580..122M,2004MNRAS.348..866E}. 
The apparent magnitude cutoff of a survey causes sparser galaxy number density at higher redshift. 
In order to form galaxy groups of similar over-density regions throughout the survey, the linking length should take into account the varying absolute magnitude cutoffs at different redshifts.
We take the standard Schechter luminosity function, $\phi(M_{R_c})$, with a luminosity evolution approximated as $M_{R_c}(z)=M_{R_c}(0)-zQ$, where Q=1.24 for red galaxies and Q=0.11 for blue galaxies \citep{1999ApJ...518..533L}. 
We adopt $M_{R_c}^*=-21.41$ and the faint end slope $\alpha=-1.20$ \citep{1997A&A...320...41K}. 
The linking length is scaled as:
        \begin{displaymath}
        D0 \propto R_w^{1/2},
        \end{displaymath}
where 
	\begin{equation} \label{Rw}
	R_w = \left(\frac{\int^{M_{cut}}_{-\infty} \Phi(M_{R_c}) dM_{R_c}}{\int^{M_{lim}}_{-\infty} \Phi(M_{R_c}) dM_{R_c}}\right).
	\end{equation}
In Equation \ref{Rw}, $M_{cut}$ is the desired absolute magnitude depth and $M_{lim}$ is the absolute magnitude limit corresponding to the apparent magnitude limit of the survey. 
This scaling factor $R_w$ is unity if $M_{cut} < M_{lim}$. 

We increase the linking length by $\sqrt{\frac{\sum^N_i w_i}{N}}$ to conserve the local galaxy number density due to the removal of unreliable galaxies, where \it N \rm is the total number of galaxies joined into a group and $w_i$ is the completeness weight (described in $\S$ \ref{sec_sample}) of each linked galaxy.  

In practice, our linking length used to search for connected galaxies in the transverse direction is expressed as:
        \begin{equation} \label{D0}
        D0=\sqrt{\frac{R_w\sum^N_i w_i}{N}}\frac{D0_{xy}}{1+z}.
        \end{equation}

\subsection{The Redshift Friendship Criterion, $P_{ratio,crit}$ } 
In the ideal situation where there is no uncertainty in the redshift, the occurrence of a galaxy or group at its redshift is a $\delta$ function. 
From a statistical viewpoint, the occurrence of an event in photometric redshift space for each galaxy is independent in the sense that the photometric redshift of each galaxy is estimated by applying a set of solutions from an empirical method. 
Given that galaxy A, galaxy B, ... , and galaxy \it n \rm with photometric redshift probability density $P_A(z), P_B(z)$, ... , and $P_n(z)$ form a group in redshift, the group redshift density is the likelihood for all these \it n \rm members to occur at the same redshift:
        \begin{displaymath}
	P_{group}(z)=P_A(z)P_B(z)...P_n(z).
	\end{displaymath}
Therefore, the main idea of our group-finding algorithm is to narrow down the photometric redshift uncertainty of a group by way of joining new galaxy members, because the group redshift is where all members in the same group may occur. 

\indent  Whether a galaxy is in the same redshift space as another galaxy is determined by the overlapping probability based on their photometric redshift probability densities. 
We use a probability ratio, $P_{ratio}$, as the criterion to set the membership in redshift. The $P_{ratio}$ for galaxy $i$ with respect to the group redshift density is defined as 
	\begin{displaymath}
	P_{ratio} = \frac{\int_{0}^{\infty} P_i(z) P_{group}(z) dz}{maxP}. 
	\end{displaymath}
The numerator is the total probability density for galaxies to occur at the same redshift. 
The denominator is the maximum value of the numerator, which occurs when all the galaxies are at the same redshift. 
To clarify the $P_{ratio}$ concept we assume two galaxies with Gaussian photometric redshift probability density $P_{z_1,\sigma_1}(z)$ and $P_{z_2,\sigma_2}(z)$, where $z_1$ and $z_2$ are the photometric redshifts for these two galaxies and $\sigma_1$ and $\sigma_2$ are the uncertainties. 
The total probability for the galaxies to occur at the same redshift is:
	\begin{displaymath}
	P = \int_{0}^{\infty} P_{z_1,\sigma_1}(z') P_{z_2,\sigma_2}(z') dz'.  
	\end{displaymath}
The maximum total probability, \it maxP\rm, occurs when $z_1 = z_2$.

	We limit these two galaxies so that they must have $z_1$ and $z_2$ separated by less than $\sigma_1+\sigma_2$. 
Based on this qualification of friendship, the extreme case is when $z_2$ is $\sigma_1+\sigma_2$ apart from $z_1$. It is worth noting that the total probability is immaterial as the friendship guideline, for the reason that this total probability depends on the standard deviations of the two photometric redshift probability density functions. 
We find that $P_{ratio}$ ranges from $\sim 0.37$ for two Gaussian probability densities of $\sigma_1 = \sigma_2$ and $|z_2 - z_1|=\sigma_1 + \sigma_2$, to $\sim 0.50$ when one of the $\sigma$ is small relative to the other. 
We set a criterion, $P_{ratio,crit}$, as the friendship criterion in redshift. 
For any galaxies to be joined together, they must have their $P_{ratio} \geq P_{ratio, crit}$.

\section{The Probability Friends-of-Friends Algorithm \label{sec_pFoF}}
\subsection{The Algorithm}
The algorithm starts with a seed galaxy, and treats every galaxy in the sample as a seed. Steps to form a group are as follows. 

\noindent \bf Step 1: The seed galaxy \rm \\
\indent $\bullet$ A 2D linking length is calculated based on this seed galaxy's photometric redshift and completeness weight (Equation \ref{D0}). \\ 
\indent $\bullet$ Galaxies within this length from the seed are searched in the transverse direction.  \\
\indent $\bullet$ Among those galaxies encircled by the linking length, a galaxy is chosen as the seed's companion which have the maximum $P_{ratio}$ relative to the seed galaxy and satisfies the condition $P_{ratio} \geq P_{ratio,crit}$. 

\noindent \bf Step 2: The proto-group \rm \\
\indent $\bullet$ The seed and its companion form a proto-group. \\
\indent $\bullet$ Calculate the photometric redshift probability density $P_{group}$(z) for the proto-group. \\
\indent $\bullet$ Assign the peak of $P_{group}(z)$ as the redshift of this proto-group.  \\
\indent $\bullet$ Re-calculate the linking length and absolute magnitudes for these two galaxies based on the assigned proto-group redshift. \\
\indent $\bullet$ Re-verify the membership of the companion by checking that:  \\
	\indent (1) the companion is still enclosed by the updated linking length; \\
	\indent (2) the companion still has the maximum $P_{ratio}$ to the seed; and \\
	\indent (3) the revised absolute magnitudes for both the seed and the companion still satisfy the sample depth $M_{R_c,cut}^{k,e}$ criterion. \\
\indent $\bullet$ A proto-group is confirmed if the membership is verified.

\noindent \bf Step 3: The primary group \rm \\
\indent $\bullet$ Examine other galaxies located within the linking length to the seed galaxy using the redshift and linking length based on the proto-group. \\
\indent $\bullet$ From these remaining galaxies, a new member is chosen which satisfies the $P_{ratio} \geq P_{ratio,crit}$ condition, and also has the highest $P_{ratio}$ to the group photometric redshift probability density. \\
\indent $\bullet$ Re-calculate the group photometric redshift probability density and the linking length with the new member included. \\
\indent $\bullet$ Re-compute the absolute magnitude of each linked galaxies using the updated group redshift.  \\
\indent $\bullet$ Re-check the membership of all connected galaxies by the $D0$ and $M_{R_c,cut}^{k,e}$ criteria. \\
\indent $\bullet$ Repeat the procedure until all the galaxies enclosed by the seed galaxy's linking length have been examined. \\
\indent $\bullet$ A primary group is formed. 

\noindent \bf Step 4: The friends-of-friends \rm \\
\indent $\bullet$ A new member is selected using a procedure similar to Step 1 in choosing the companion, but applied to galaxies within the linking length of any members in this primary group. \\
\indent $\bullet$ Repeat the process for all members of the primary group until there are no more additional galaxies linked or rejected. \\
\indent $\bullet$ A `mini-group' is formed. The prefix `mini-' refers to the group associated with each seed galaxy. 

\noindent \bf Step 5: The mini-groups \rm \\
\indent $\bullet$ Steps 1 to 4 are carried out for all galaxies.  
Since each galaxy in our sample is considered as a seed galaxy, each galaxy has its own mini-group. 

\noindent \bf Step 6: Unifying mini-groups \rm \\
\indent The procedure of unifying mini-groups is necessary since a galaxy may be a member of many mini-groups.
The unifying principles are similar to those used to form mini-groups; that is, mini-groups must have some common members and satisfy a $P_{ratio,crit}$ threshold in order for them to merge into a more massive group. Terminologically, we refer to the mini-group formed using seed galaxy $X$ as `mini-group $X$'. 
We detail the procedures below, with mini-group $A$ having $N$-1 other members $X_i$, where $i=1$ to $N-1$. \\
\indent $\bullet$ If the photometric redshift probability density of mini-group $X_i$ satisfies the $P_{ratio,crit}$ criterion with respect to that of mini-group $A$, all members of mini-group $X_i$ are added to the member list of mini-group $A$; otherwise, galaxy $X_i$ will be removed from the member list of mini-group $A$. 
The addition and removal of galaxies from mini-group $A$ takes place only after all mini-groups $X_i$ have been checked. \\
\indent $\bullet$ Since the process of merging or removal will affect the redshift probability density of mini-group $A$ and hence may fragment the mini-group, the following criteria must all be satisfied individually for a surviving member $X_i$ and its mini-group members after the merging process above: \\
\indent  (1) the member satisfies the $P_{ratio,crit}$ to the updated mini-group $A$ probability density; \\
\indent  (2) the member has at least one member of mini-group $A$ within the linking length; and \\
\indent  (3) the member is still brighter than $M_{R_c,cut}^{k,e}$ at the updated group redshift.

In some circumstances, an original member of a mini-group may have already been flagged as belonging to other merged group(s). For instance, the member list of mini-group $A$ is 
	mini-group $A$ = \{A, n2, n3, n4, ... , n8, n9\}, 
where n2, n3, n4 also belong to `grp\#1', while n6 and n7 are members of `grp\#2'. The subsequent classification of mini-group $A$'s members may belong to one of the following cases:

\indent (1) If all mini-group $A$'s members have their $P_{ratio}$ satisfying the $P_{ratio,crit}$ criterion to all overlapping groups (i.e., `grp\#1' and `grp\#2'), the member lists of mini-group $A$ and the overlapping groups are merged together and all these groups share the same group ID.
In other words, mini-group $A$ has the role as being a `bridge' in connecting these overlapping merged groups. \\
\indent (2) If some of the mini-group $A$'s members have $P_{ratio} \geq P_{ratio,crit}$ to an overlapping group (e.g., `grp\#1') and some other mini-group $A$'s members satisfy the $P_{ratio,crit}$ criterion to another overlapping group (e.g., `grp\#2'), the member list of mini-group $A$ is delisted and all its members are classified into these overlapping groups. For the situation that some of the mini-group $A$'s members satisfy the $P_{ratio,crit}$ criterion to more than one overlapping groups, these members are classified into the overlapping group of the best $P_{ratio}$.
 
\indent $\bullet$ After every mini-group has been examined, a final group catalog is established.

\subsection{Discussion}
We name our group-finding algorithm \it `Probability Friends-of-Friends' \rm (pFoF) for its two main characteristics of group redshift probability density and the FoF 2D linking. 
The main feature in our group finding procedure is the dynamic linking. 
The group redshift probability density and the linking length keep being refined through the entire process and are used to re-check all connected galaxies in this group. Some interesting points are:

\indent  (a) the normalized group redshift probability density is reconstructed every time when a galaxy joins to or is rejected from this group as
	\begin{displaymath}
	P_{group}(z) = \frac{P_1(z) P_2(z) ... P_n(z)}{\int P_1(z') P_2(z') ... P_n(z')dz'}.
	\end{displaymath}
\indent (b) The use of the above group redshift probability density in calculating $P_{ratio}$ for a new galaxy can be interpreted as the probability for this new galaxy to be in this group, given N members at the same redshift.\\
\indent  (c) The absolute magnitude of the connected members are re-computed, and the members are re-checked using the updated linking length every time when any galaxy is connected or rejected. \\
\indent  (d) Algorithmically, a single galaxy is considered as a group as well. 
In subsequent analysis, we set a minimum of five galaxies in a group to exclude groups with too few galaxies, so that group redshift can be well confined by its members.

  One different approach in applying this `photometric redshift probability density' idea in group finding, in place of mini-groups and the unifying procedure, is to continue Step 4 until no more new members are linked. 
However, we find that this alternative group finding procedure may break a massive group (usually, a galaxy cluster) into several pieces in redshift space, especially in the region where the galaxy number density is extremely high, such as the core of a cluster. 
This happens because the formation of a massive galaxy aggregation has confined the group redshift to be in a narrow redshift space, and gives no flexibility for other galaxies of sufficiently different photometric redshifts to join in. 
These `other galaxies' are usually the outliers in the comparison of the photometric and spectroscopic redshifts for individual galaxies. 
The idea of unifying mini-groups reduces the degree of the splitting of massive galaxy aggregations, but this still cannot be absolutely avoided unless higher accuracy photometric redshift measurements are available.

   In carrying out the group finding, we sort the sample galaxies by their peak values of the photometric redshift probability densities. 
The role of galaxy orders mainly lies in the steps of unifying `mini-groups', where the existing `mini-groups' (or merged ones) are used to combine with more `mini-groups' with lower ranks. Using mock catalogs (see $\S$\ref{sec_test}), we have tested the effect of the ordering of the seed galaxies and found that it has negligible influence on the results. 
We still decide to sort our catalogs by the peak values of the photometric redshift probability, so that each final group grows from the `mini-groups' of seed galaxies with the best quality.

\section{The pFoF Group Properties \label{sec_grp}}
\subsection{Group richness}
\indent We use $N_{gz}$ to denote the number of linked galaxies. 
The group richness, $N_{gal}$, is indicated by the total completeness weight $w_i$ for galaxies in the group with background galaxy counts subtracted: 
	\begin{displaymath}
	N_{gal}= R_w \sum_{i}^{N_{gz}} w_i - A_{grp}\Sigma_{grp,bg},
	\end{displaymath}
where $A_{grp}$ is the group area and $\Sigma_{grp,bg}$ is the contaminating background galaxy surface density within the group. 
These two quantities are estimated from the data as described 
in the following two subsections. 
In other words, $N_{gal}$ is the net number of members in a pFoF group. 
We select pFoF groups which contain at least five physically linked galaxies (i.e., $N_{gz} \geq 5$) so that the group redshift can be well restricted by the members. 

\subsection{Background galaxy density in galaxy groups}
\indent The background galaxy surface density is estimated from the complete photometric redshift catalogs; in our case, the RCS1 CFHT patches (Hsieh et al. 2005). 
We apply the same cutoffs both in magnitude and photometric redshift as our galaxy sample selection. The completeness weight for each galaxy is considered as well. 
We then calculate the number of background galaxies per Mpc$^2$ in photometric redshift bins of 0.01, and express it as $\Sigma_{bg}(z)$. 
This $\Sigma_{bg}(z)$ has taken the scaling factor $R_w$ (Equation \ref{Rw}) into consideration. 

The pFoF algorithm allows us to constrain group redshift within $\Delta z_{grp}< 0.02$ although photometric redshift uncertainties of member galaxies can be as large as $\sigma_{emp}\sim 0.070$.
Therefore, to estimate the background galaxy contamination within a galaxy group, we should consider the photometric redshift space within which all members of a group may occur, i.e., the likelihood. 
Accordingly, to form the likelihood, we sum the photometric redshift probability densities of all members and normalize the peak of this summed photometric redshift distribution to unity, denoted as $L(z)$. 
The background galaxy density for this group is estimated using this photometric redshift likelihood as 
	\begin{equation}
	\Sigma_{grp,bg}=\int_0^{\infty} L(z)\Sigma_{bg}(z)dz. \label{grpbk}
	\end{equation}
The $L(z)$ has broader wings and wider width than $P_{group}(z)$. The estimation of $\Sigma_{grp,bg}$ is underestimated if $P_{group}(z)$ is used instead in Equation \ref{grpbk}.
This is because the $L(z)$ represents the redshift that a galaxy in a group could have if we drew it from that group.

\subsection{Projected group area}
\indent Geometrically, the mean separation among $N$ galaxies uniformly distributed over an area $A$ is 
	\begin{displaymath}
	<s> = \sqrt{A/N}.
	\end{displaymath}
If we assign each galaxy a circular area of radius $r$, we should expect the total area of these circles centered at individual galaxies to be the same as the total area $A$, i.e.:
	\begin{displaymath}
	N \pi (k <s>)^2 = A, 
	\end{displaymath}
where $r = k<s>$. Consequently,
	\begin{displaymath}
	k = \frac{1}{<s>} \sqrt{A/N} = \frac{1}{\sqrt{\pi}}.
	\end{displaymath}
We calculate the projected group area using an empirical method. 
Each member in a group is assigned a radius $r=<s>/\sqrt{\pi}$, where $<s>$ is computed as $1/\sqrt{\Sigma_{grp,bg}}$. 
We then draw a rectangular box of the area $A_{reg}$ with the length and width enclosing the R.A. and Dec. range of the circles centered at each group member. $N$ random uniformly distributed points are casted over this rectangular box. 
By counting the numbers ($N_{in}$) of these $N$ points within the distance $r$ to any group member, the projected group area is computed as 
	\begin{displaymath}
	A_{grp}'=\frac{N_{in}}{N}A_{reg}.
	\end{displaymath}
Consequently, the estimated background galaxy number in a pFoF group is calculated as $N_{bg}'=A_{grp}'\Sigma_{grp,bg}$. 
However, since galaxies are not distributed uniformly, this background estimation must be considered as a lower limit. 
Tests performed on mock catalogs allow us to cross-check the true and computed contaminating background galaxy counts within a galaxy group. 
From these tests, we find that the computed background galaxy counts in a pFoF group are correlated with the true number of contaminating galaxies, but not with a unity slope (see \S7.2.1).
Hence, equivalently, we can apply an empirical correction to the projected group area to obtain an effective area, so that the background galaxy counts are properly estimated:
        \begin{equation}
        A_{grp}= 1.634 \frac{N_{in}}{N}A_{reg} - \frac{2.505}{\Sigma_{grp,bg}} \label{eq:afactor},
        \end{equation}
based on the results from simulated catalogs.
We note that the empirical corrections are similar (within 10\%) for a variety of linking criteria and sample selections.

\section{Testing pFoF Algorithm on Mock Catalogs \label{sec_test}}
\subsection{Mock Catalogs \label{sec_mock}}
To assess the quality of the pFoF algorithm, we perform tests using mock catalogs which have been obtained by the Virgo Consortium Millennium Simulation \citep{2005Natur.435..629S} using semi-analytical modeling of galaxy evolution by \cite{2006MNRAS.365...11C}.
Groups in the simulation are identified by a FoF group-finder with a linking length of 0.2 of the mean particle separation \citep{2006MNRAS.365...11C}. 
We prune off those FoF halos which contain only one or two galaxies, and define galaxies in these poor FoF halos as field galaxies.

  Our mock catalogs contain $\sim 800,000$ galaxies in $BRI$ magnitudes with $R_{AB}<26.0$ with redshifts extending from 0 to 1.4 in a total of 5.0 square degrees of sky area from five cones. 
For the purpose of testing our algorithm, we convert the photometry in the mock catalogs to the Vega system, and set a cutoff as $m_{R_c} < 22.5$ to mimic a flux limited sample. 
With this apparent magnitude cutoff, the sample becomes incomplete at $M_{R_c,cut}^{k,e}=M_{R_c}^*+2.0$ at $z_{cut} = 0.412$. 
To simulate photometric redshift for the total of 177,344 galaxies in our mock sample, we take the following steps.

\indent $\bullet$ We construct photometric-redshift functions using our training set galaxies in each spectroscopic redshift bin with size of 0.05.
The histogram of the computed photometric redshifts of these galaxies in each bin is normalized to have an area of unity, which forms the photometric redshift distribution function for that redshift bin. \\
\indent $\bullet$ The photometric-redshift distribution functions are then used to draw a photometric redshift for each galaxy in the mock sample in the corresponding redshift bin, so that any offset between photometric and spectroscopic redshifts in the real observational samples can be mimicked. The use of the photometric-redshift distribution function derived from the actual sample also ensures that the dispersion between the simulated photometric and true redshifts increases toward higher redshifts. \\
\indent $\bullet$  Each galaxy in the mock sample is then tagged with a photometric redshift probability density centered at its simulated photometric redshift. 
The tagged photometric redshift probability density is based on that associated with a training set galaxy of similar color and magnitude. 
This enables us to obtain reasonable dependence of photometric redshift probability density on galaxy color and magnitude, so that the distributions of photometric redshift uncertainties for galaxies in the mock sample are similar to those of our training set galaxies. 

The dispersion between the simulated photometric redshift and actual galaxy redshift for galaxies in the mock sample is $\sim 0.061$ at $0.3 \leq z_{mock} < 0.6$, and $\sim 0.122$ at $0.6 \leq z_{mock} < 0.9$ (compared with 0.060 and 0.134 in our real data set). 

  After the simulated photometric redshifts are obtained, we carry out the sample selection criteria for those galaxies in the mock sample. 
The completeness factor $w_i$ is computed and assigned to each galaxy satisfying the selection. 
We find that $w_i \sim 1.29$ at $m_{R_c}=22.5$. 
We also select galaxies in the mock catalogs brighter than $M_{R_c}^*+2$ after applying approximate k- and evolution corrections. 
A total of 72,954 galaxies are in our final selected mock sample, and the median $w_i$ is $\sim 1.09$. 
We refer to this simulated photometric redshift sample resembling our real data as the `$z_{simulated}$' sample.

\subsection{Test Results \label{res_test}}
  We apply our pFoF group-finding algorithm to the mock photometric redshift sample with fiducial parameters $P_{ratio,crit}=0.37$ and $D0_{xy}=0.25$ Mpc. 
We use the mock photometric redshift sample itself as the control field for background subtraction. 

\subsubsection{contaminating background galaxies}
   Background galaxy contamination correction is essential for any work using photometric redshifts. 
The photometric redshift technique can be an effective tool in scientific analysis, if the estimated and true background galaxy contamination are comparable to each other. 
For each pFoF group, we estimate the numbers of background galaxies as $N_{bg}' = A_{grp}'\Sigma_{grp,bg}$ as described in $\S$\ref{sec_grp}. 
In the use of mock catalogs, we can count the actual contaminating galaxies; i.e., $N_{bg,actual}$, galaxies contributed by the field, or other halos, or both. 
By comparing $N_{bg}'$ and $N_{bg,actual}$ in each true pFoF group, we find that $N_{bg}'$ tends to be underestimated when $N_{bg,actual}$ is large and the trend can be approximated using a linear relation as $N_{bg,actual} = 1.634 \times N_{bg}' -2.505$. 
We therefore apply the linear relation to correct $N_{bg}'$ by adjusting the group area $A_{grp}'$ (Equation \ref{eq:afactor}). 
We use $N_{bg}$ to denote the number of the estimated background galaxies with the linear correction applied.

\subsubsection{Test 1: the recovery rate \label{mock:recovery}} 
    To test the performance of our pFoF algorithm, we first investigate the group recovery rate of the mock sample. 
We apply our pFoF group-finding algorithm to the mock $z_{simulated}$ sample with $P_{ratio,crit}=0.37$ and $D0_{xy}=0.25$ Mpc. 
The mock groups which have at least three members brighter than our sample magnitude cutoffs (i.e., $m_{R_c} < 22.5$ and $M^{k,e}_{R_c} < M^*_{R_c}+2.0$) are selected as the reference groups, with a total number of 705 at $z < z_{cut}$. 

We use the following matching procedure. Since every galaxy has a pFoF group ID in the output files of the pFoF algorithm, we classify each member of a mock reference group by its pFoF group ID. 
The members of a given mock group may belong to different pFoF groups. 
We define the pFoF group which matches the mock group as the one that contains the largest number of members of the mock group and also satisfies $N_{gal} \geq 3$ and $N_{gz} \geq 5$. 
Each pFoF group is allowed to match with only one reference mock group. 
If there is more than one reference mock group recovered by the same pFoF group, only one of these reference mock groups is classified as `recovered'. 

\begin{figure}
        \includegraphics[width=12.7cm]{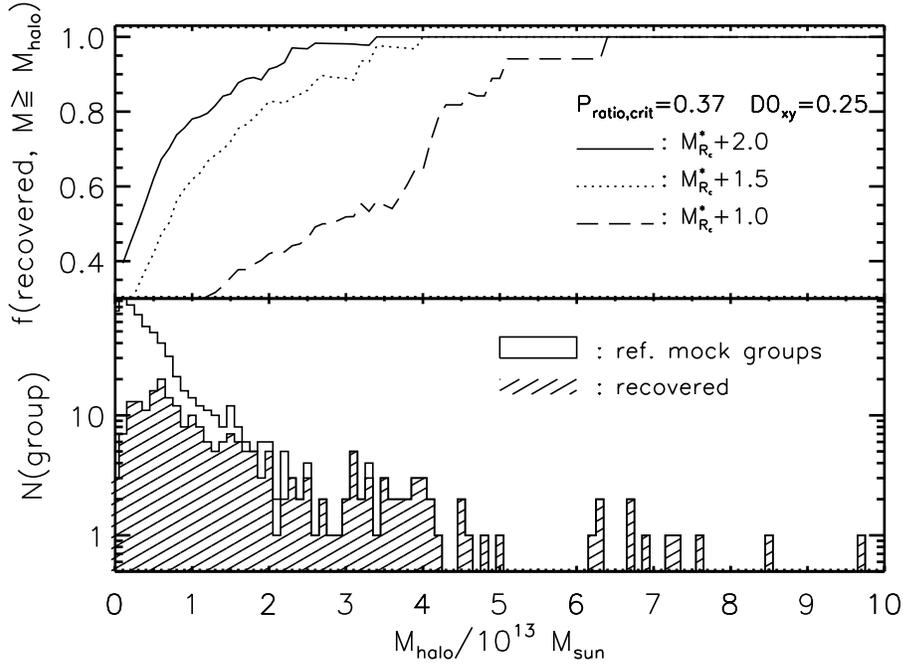}        
	\caption[Test 1 \& Test 3: the recovery rate]{The results of Test 1 using pFoF groups obtained with $P_{ratio,crit}=0.37$ and $D0_{xy}=0.25$ Mpc. The recovery rates as a function of mock group halo mass are shown in the top panel. The three curves represent the results using three different sample depths as indicated in the panel. The distribution of the reference mock group halo mass with the $M_{R_c}^*+2.0$ cutoff is shown as the un-hatched histogram in the bottom panel, and the recovered reference groups are presented as the hatched histogram. \label{mockplot}} 
	\end{figure}

The results of the recovery test are presented in Fig. \ref{mockplot}. 
The Y-axis in Fig. \ref{mockplot} is the fraction of the recovered to the total reference mock groups of halo mass greater than a cutoff (i.e., the X-axis). 
The recovery rate increases when the halo mass is larger. 
The pFoF algorithm recovers more than 80\% of the reference mock groups of halo mass greater than $\sim 1.2\times 10^{13}M_{\sun}$, and recovers all mock groups of halo mass greater than $\sim 3.4 \times 10^{13}M_{\sun}$. 
The total number of reference mock groups with mass larger than the two above mentioned limits are 147 and 41, respectively.
The r.m.s. dispersion in redshift between the recovered reference mock groups and the matched pFoF groups is $\sim0.044$, and it is improved to $\sim0.038$ for groups with halo mass greater than $\sim 3.4 \times 10^{13}M_{\sun}$. 

\subsubsection{Test 2: the fractions of false detections and serious projections}
\indent  To investigate the fraction of false pFoF groups, we examine every member of a pFoF group to see in which mock halos they are located. 
With $P_{ratio,crit}=0.37$ and $D0_{xy}=0.25$ Mpc, we have a total of 1,019 pFoF groups as the reference, selected with $N_{gal} \geq 3$, $N_{gz} \geq 5$, and $z_{pFoF} < z_{cut}$. 
A pFoF group is flagged as `false detection' if either: \\
\indent (1) all its members are composed of field galaxies (i.e., galaxies in poor mock halos containing fewer than three galaxies), or\\
\indent (2) it contains fewer than three members from the mock group with the largest matched members.

\begin{figure}
        \includegraphics[width=12.7cm]{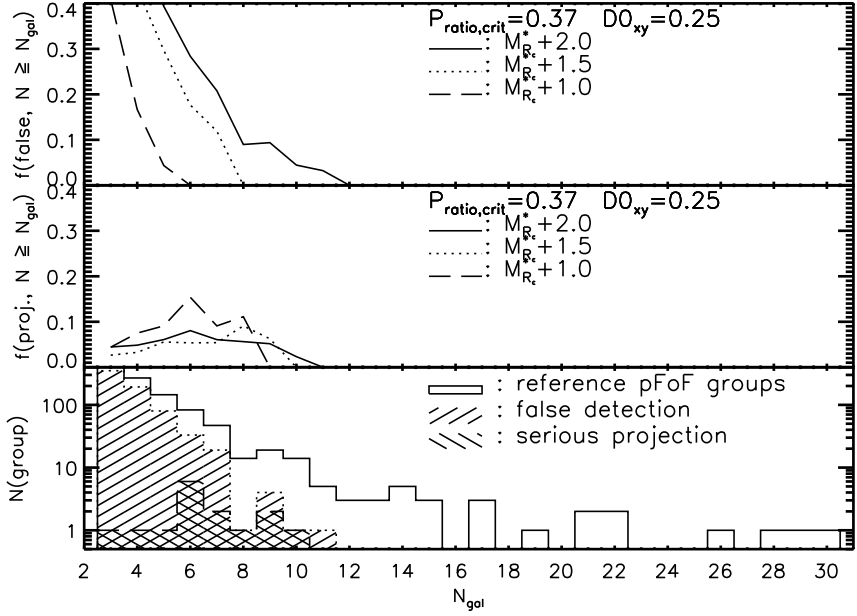}
	\caption[Test 2 \& Test 3: the rates of false detection and serious projection]{\it Top: \rm The fractions of false pFoF groups as a function of group richness for three different sample depths as indicated in the panels. \it Middle: \rm The fraction of pFoF groups flagged as 'serious projection', which are pFoF groups containing members from multiple mock groups. \it Bottom: \rm The unhatched histogram is the richness distribution for the reference pFoF groups in the sample of $M_{R_c}^*+2.0$ depth. The number of false and `serious projection' pFoF groups are presented as the hatched histograms. \label{fmockplot}}
        \end{figure}

We present the results in the top panel on Fig. \ref{fmockplot}. 
The Y-axis is the fractions of false pFoF groups (over the total) with $N_{gal}$ greater than a cutoff (the X-axis). 
The fraction of false groups decreases with increasing $N_{gal}$.
The false detection rate is 30\% for pFoF groups of $N_{gal} \geq 5.85$, and is 10\% when $N_{gal} \geq 7.91$. 
There are 222 and 79 pFoF groups of $N_{gal}$ greater than these two richness cutoffs, respectively.
We note that a pFoF group of $N_{gal} \sim 8$ corresponds to a halo mass $\sim 3.7\times 10^{13}M_{\sun}$. 
We find the fraction of false groups increases toward higher redshift. 
In these tests, all the false pFoF groups with $N_{gal}\ge 8$ are located at $z>0.34$.

A pFoF group may contain multiple mock groups if an inappropriate $P_{ratio}$ or $D0_{xy}$ is used.
To examine the fraction of such pFoF groups, we flag a pFoF group as `serious projection' if two or more mock groups contribute similar numbers of galaxies to the pFoF membership. 
Using $N_1$ and $N_2$ to denote the numbers of galaxies in a pFoF group from mock group \#1 and \#2 and $N_1 \geq N_2$, this pFoF group will be flagged as `serious projection' if $N_1/N_2 < 1.5$. 
 The results are presented in the bottom panel in Fig. \ref{fmockplot}, where the Y-axis is the fractions of `serious projection' to the total pFoF groups with $N_{gal}$ greater than a cutoff in the X-axis. The fractions of `serious projection' is about 5\% for all  $N_{gal}10$ cutoffs below $\sim$10.

\subsubsection{Test 3: the effect of magnitude limit}
\indent To test how sample depth affects the pFoF performance, we repeat Test 1 and Test 2 but with two additional different $M_{R_c,cut}^{k,e}$ cutoffs: $M_{R_c,cut}^{k,e}=M_{R_c}^*+1.0$ and $M_{R_c,cut}^{k,e}=M_{R_c}^*+1.5$. 
The results are listed in Table \ref{tab:mocktest} and overplotted in Figures \ref{mockplot} and \ref{fmockplot} as the dashed and dotted curves.

The number of recovered mock groups increases with increasing sample depth,
but the fraction of false groups increases as well when $M_{R_c,cut}^{k,e}$ changes from $M_{R_c}^*+1.0$ to $M_{R_c}^*+2.0$. 
We therefore conclude that samples with shallower depths miss a larger portion of true groups, especially the poorer ones; 
going deeper into the luminosity function increases the identification of true galaxy groups with a higher, but still acceptable, false detection rate. 
Based on these tests of different $M_{R_c,cut}^{k,e}$ cutoffs, we suggest that a sample should have a depth of at least $M_{R_c}^*+1.5$ in order to obtain better group finding results.

\subsubsection{Test 4: the linking criteria}
\begin{figure}
        \includegraphics[width=8.2cm]{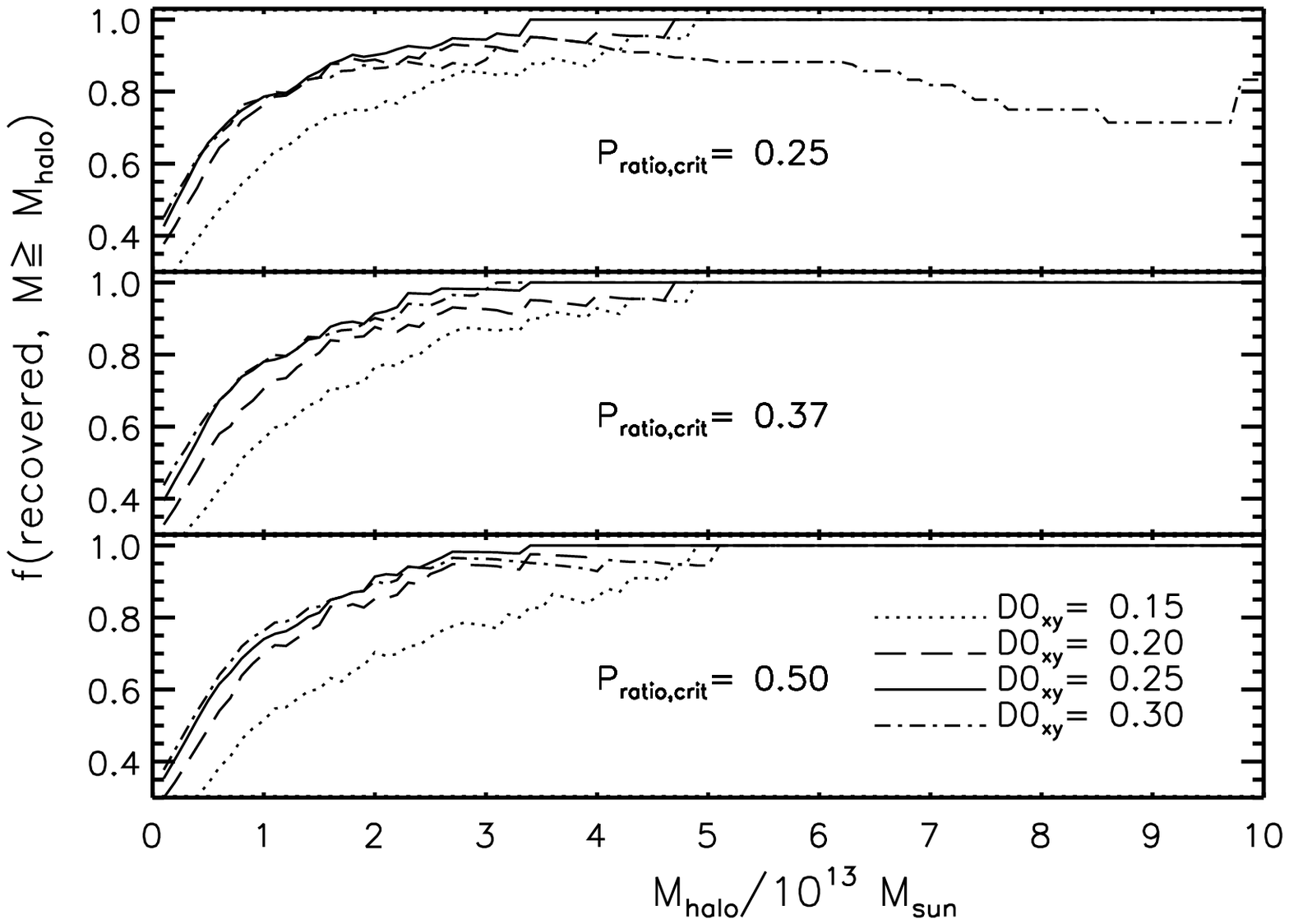}  
        \includegraphics[width=8.2cm]{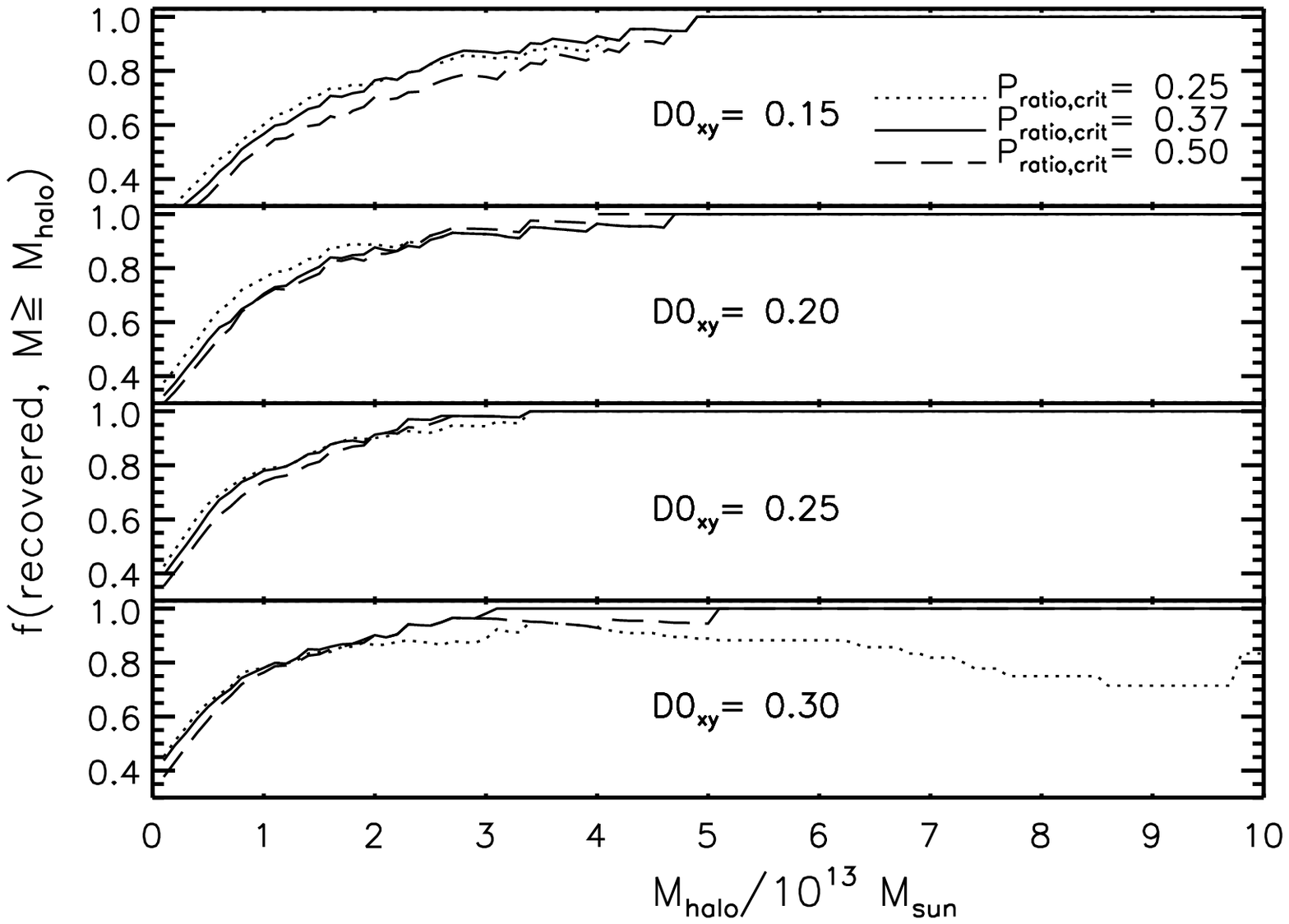}   
        \caption[Test 4: recovery rates for different linking criteria]{The recovery rate (Test 1) for pFoF groups obtained using varying $P_{ratio,crit}$ and $D0_{xy}$. The left panels plot the results for various $D0_{xy}$ for each fixed $P_{ratio,crit}$.
The right panels show the results using fixed $D0_{xy}$ with varying $P_{ratio,crit}$. 
\label{dtFOFcomp} }    
        \end{figure}

\begin{figure}
        \includegraphics[width=8.2cm]{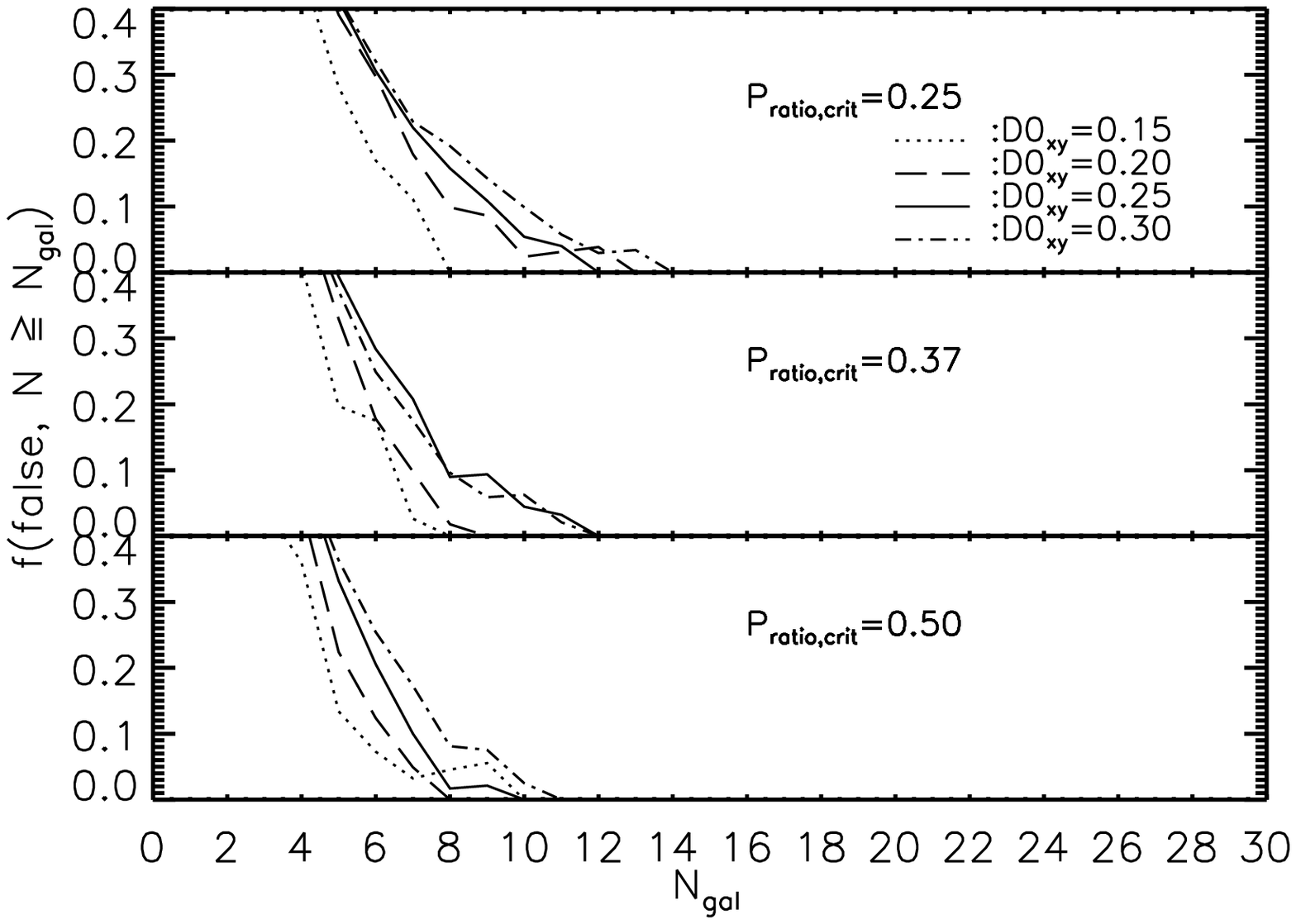} 
        \includegraphics[width=8.2cm]{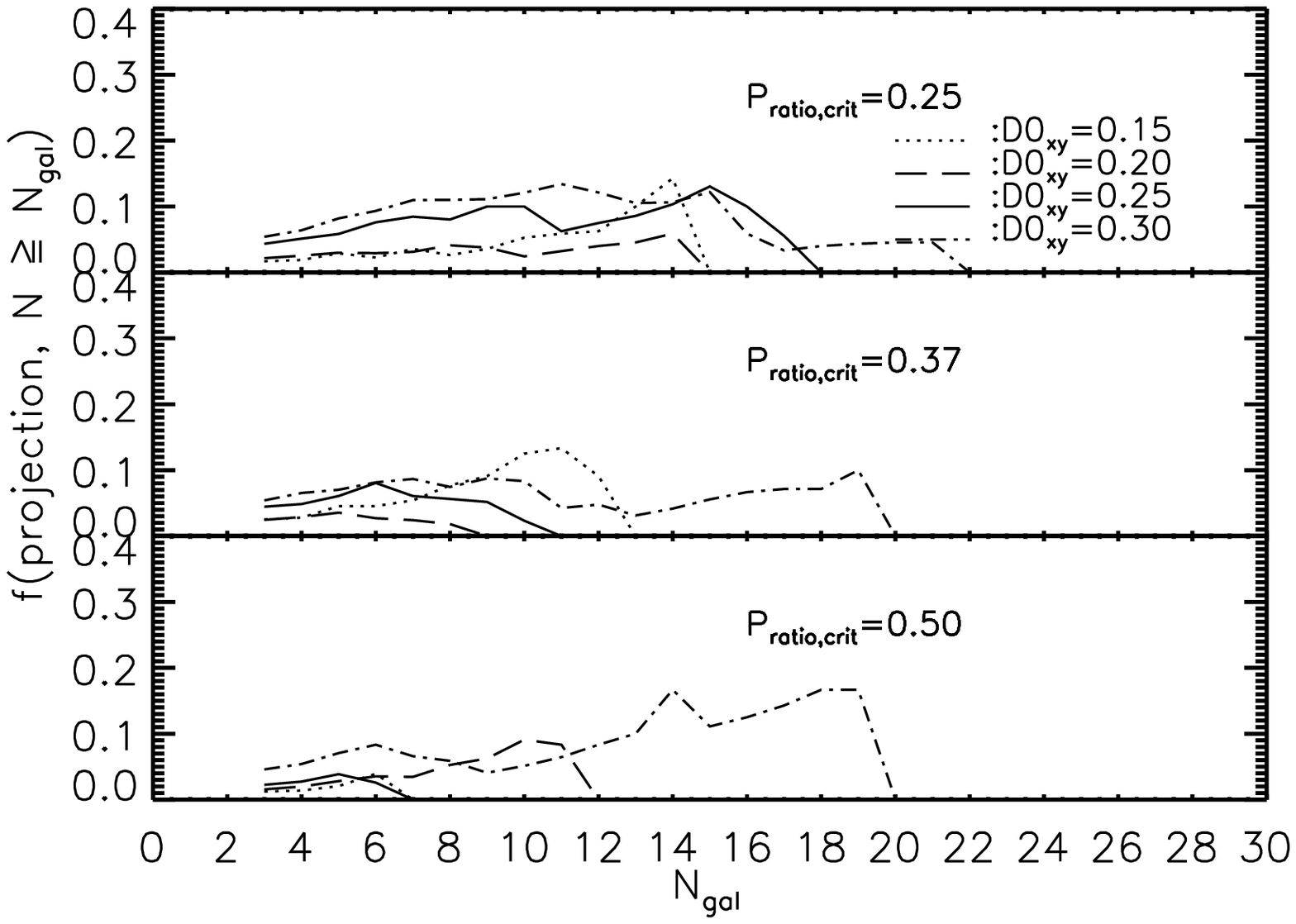}
        \caption[Test 4: false detection rate and fraction of projection using different $D0_{xy}$ and a fixed $P_{ratio,crit}$]{\it Left: \rm The false detection rates (Test 2) using various $D0_{xy}$ with a fixed $P_{ratio,crit}$. \it Right: \rm The fractions of `serious projection' (Test 2) using the same set of $D0_{xy}$ and $P_{ratio,crit}$. 
\label{dfFOFcomp}}        
	\end{figure}

\begin{figure}
        \includegraphics[width=8.2cm]{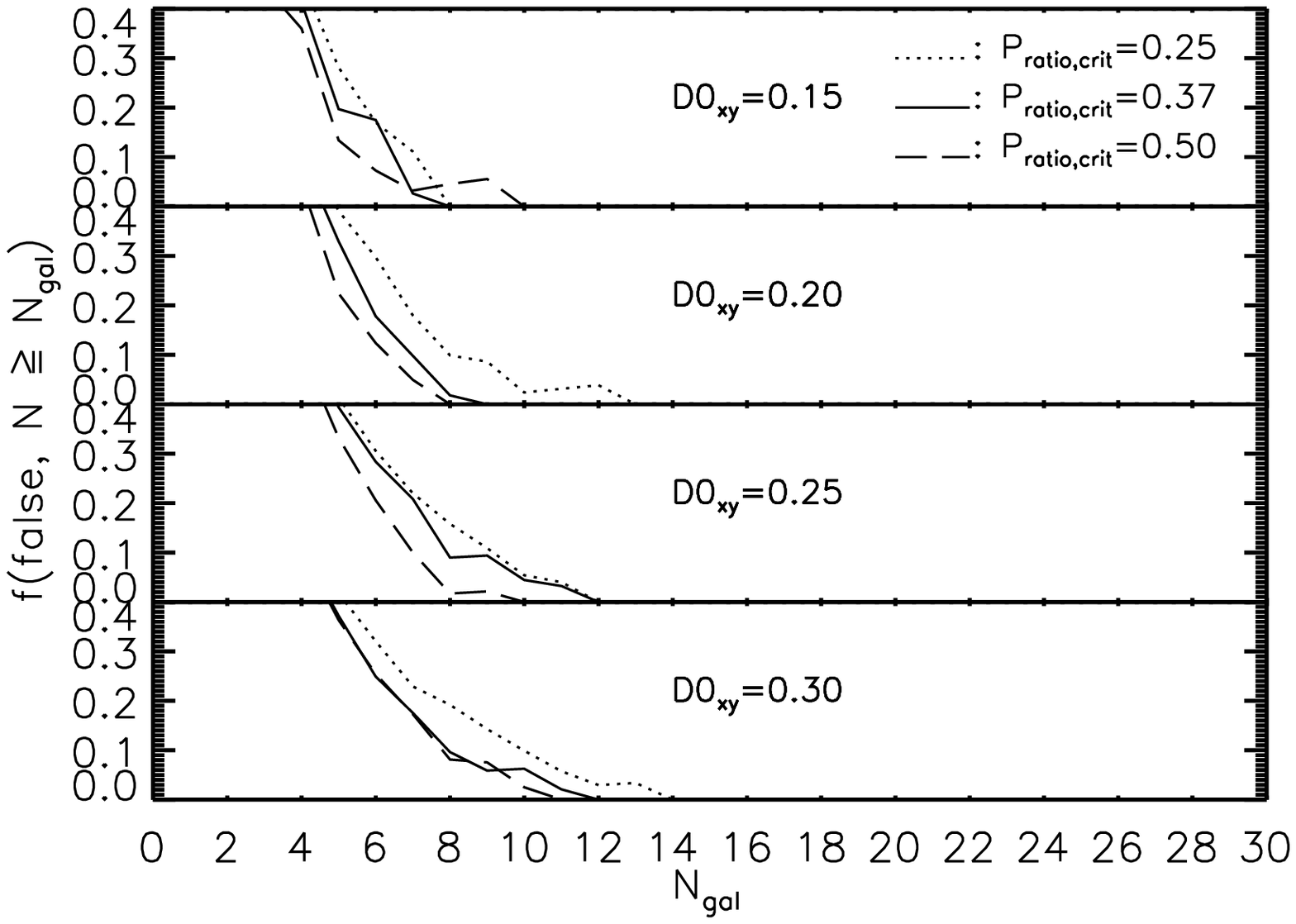}  
        \includegraphics[width=8.2cm]{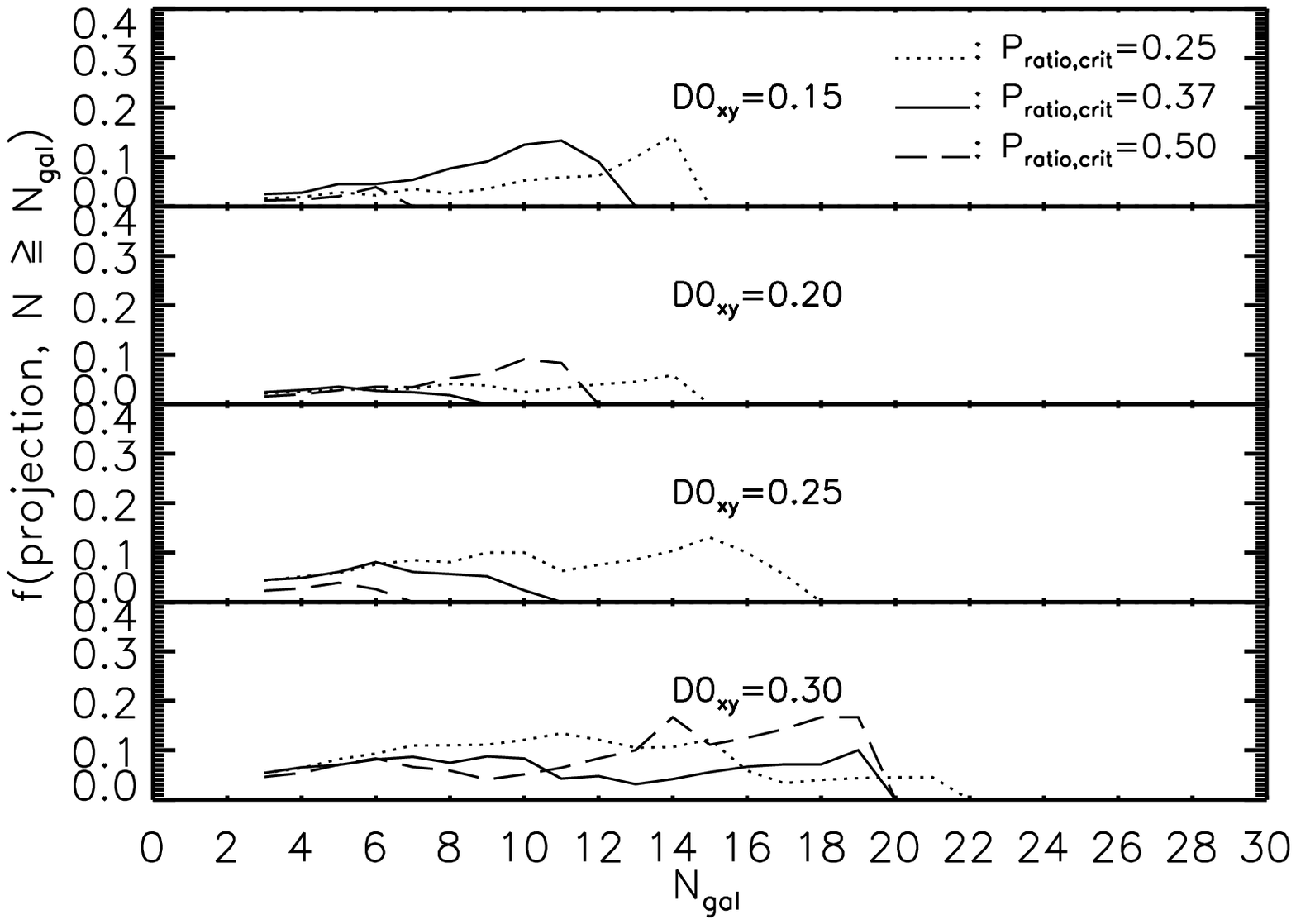}        
	\caption[Test 4: false detection rate and fraction of projection using different $P_{ratio,crit}$ and a fixed $D0_{xy}$]{Same as Fig. \ref{dfFOFcomp} but keeping $D0_{xy}$ fixed and varying $P_{ratio,crit}$. \label{pfFOFcomp}}
        \end{figure}

\indent One of the critical issues in any group-finding algorithm based on the friends-of-friends algorithm is the choice in the values of the linking parameters. 
To probe how the linking criteria affect pFoF membership, we repeat Tests 1 and 2 by changing the values of $P_{ratio,crit}$ and $D0_{xy}$. 
The results are listed in Table \ref{tab:mocktest} and presented in Figures \ref{dtFOFcomp}, \ref{dfFOFcomp}, and \ref{pfFOFcomp}.

  The tests of different linking criteria show that there is a dynamic relation between $P_{ratio,crit}$ and $D0_{xy}$. 
The use of larger linking lengths, while providing a better recovery rate, tends to form more groups which are not truly physically related. 
This higher recovery rate and larger fractions of false detection and `serious projection' groups are also applicable to tests using smaller $P_{ratio,crit}$. 
Therefore, a set of $P_{ratio,crit}$ and $D0_{xy}$ should be chosen which is a compromise between the recovery and false detection rates. 
We adopt $P_{ratio,crit}=0.37$ and $D0_{xy}$=0.25 Mpc for further tests of our algorithm. 

\subsubsection{Test 5: Gaussian probability densities}
\indent We check the performance of the pFoF algorithm under the assumption of Gaussian photometric redshift probability densities. To do this, we take each galaxy's photometric redshift and error as the mean and standard deviation to generate a Gaussian photometric redshift probability. 
We call these catalogs `Gaussian', and name `non-Gaussian' for the sample based on real photometric redshift probability densities (i.e., the `$z_{simulated}$' sample).
The completeness correction weight $w_i$ is also calculated for the `Gaussian' sample. 
The $w_i$ is $\sim 1.07$ at $m_{R_c}$=22.5, and the averaged $w_i$ is $\sim 1.03$. 
The estimated background counts are re-computed using the `Gaussian' sample, which are similar to those estimated using the `non-Gaussian' sample.

\begin{figure}
        \includegraphics[width=8.2cm]{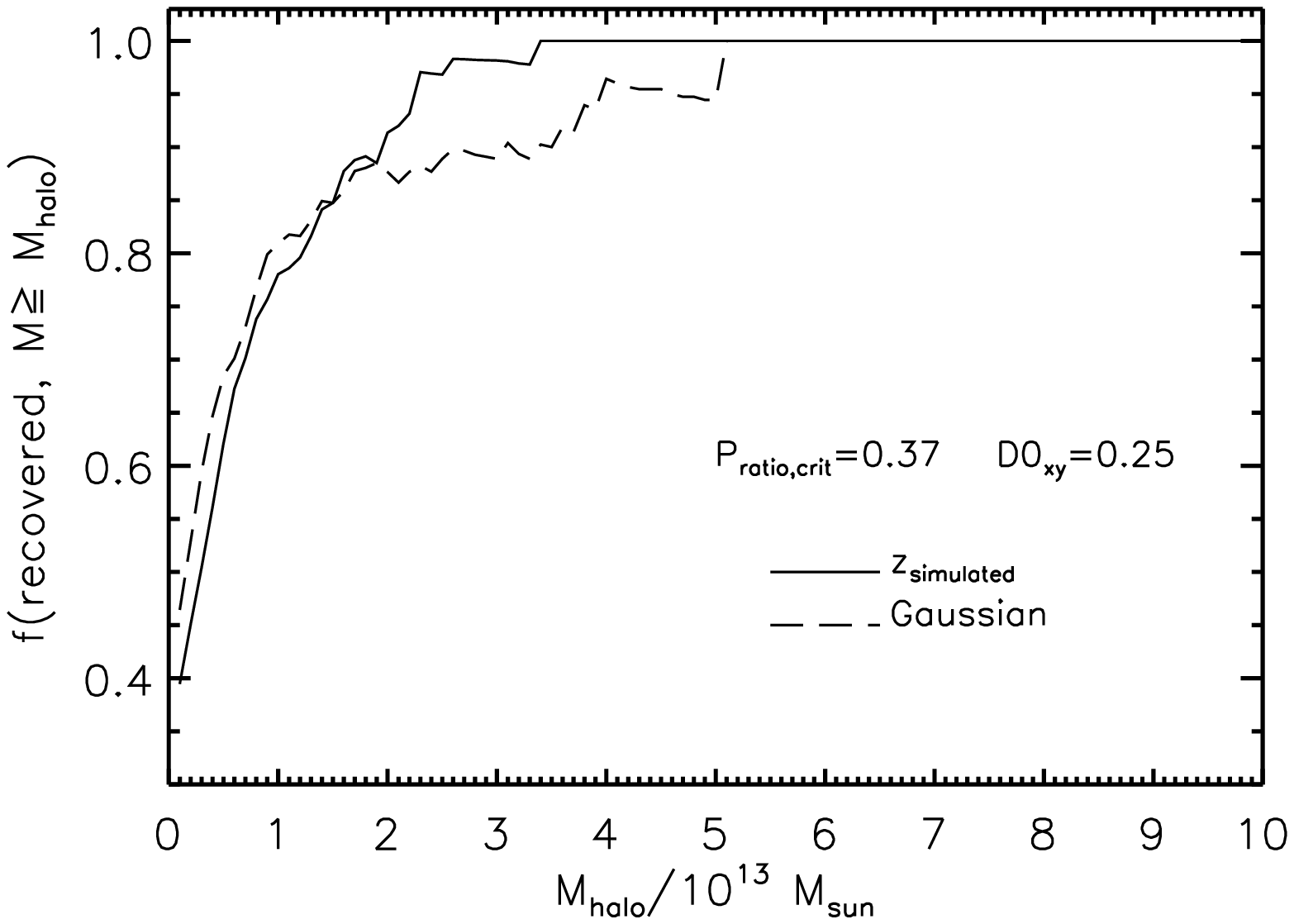}        
        \includegraphics[width=8.2cm]{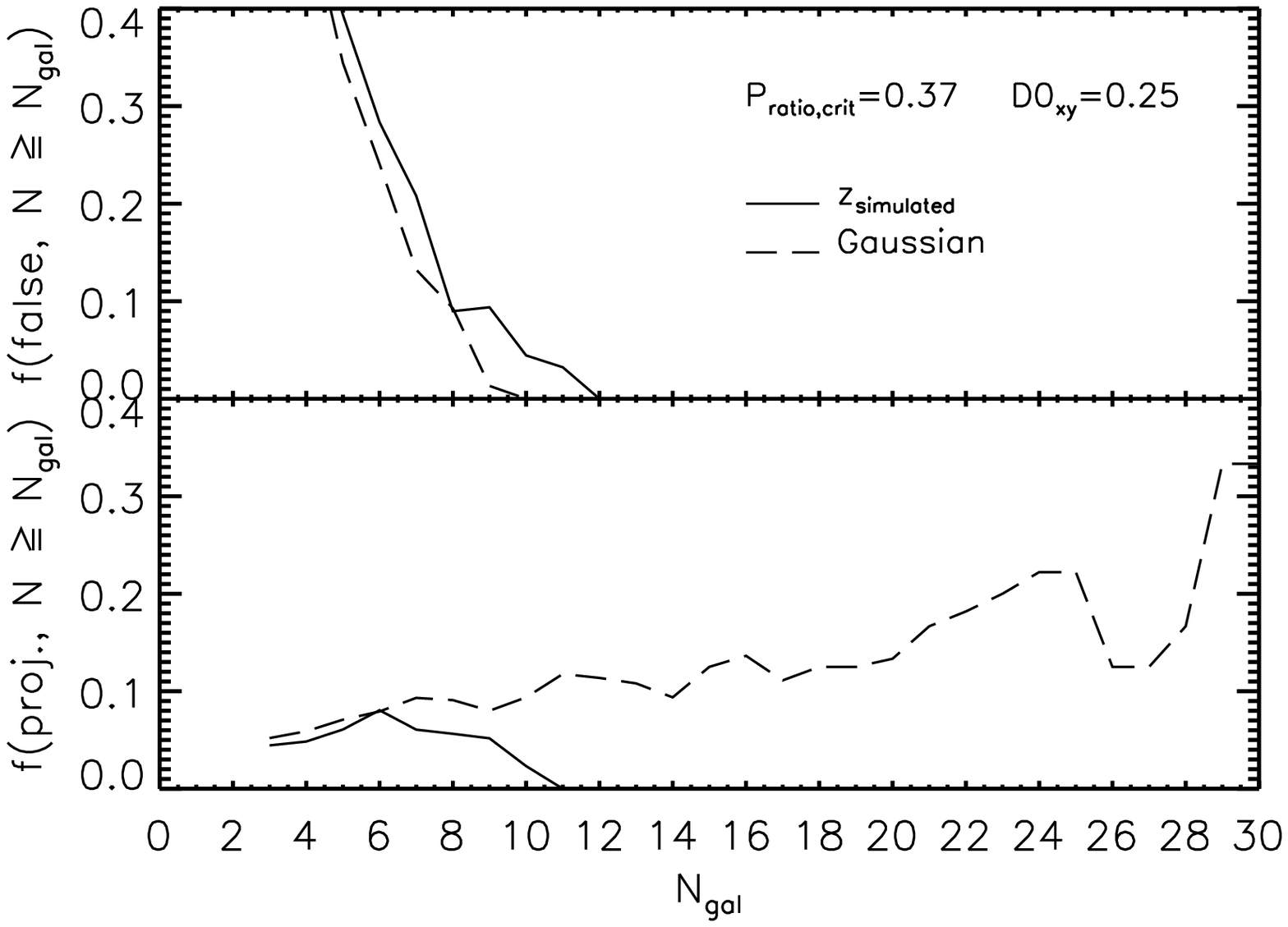}        
	\caption[Test 5: the `Gaussian' sample]{\it Left: \rm The repeated Test 1 results (recovery rate) using the `$z_{simulated}$' and `Gaussian' samples (see Test 5 in $\S$\ref{res_test}). \it Right: \rm The false detection rates (\it top\rm) and the fractions of the pFoF groups flagged as `serious projection' (\it bottom\rm) using the same two samples in the left plot. \label{gsmockplot}}
        \end{figure}

  The results of this test are illustrated as the dashed curves in Fig. \ref{gsmockplot}. 
Compared with the Test 1 results of using the `$z_{simulated}$' sample (the solid curves), the `Gaussian' sample recovers slightly more mock groups of halo mass less than $1.3 \times 10^{13} M_{\sun}$, but it fails to recover as many mock groups of halo mass $1.3-5.0 \times 10^{13} M_{\sun}$ as using the `$z_{simulated}$' sample. 
The `Gaussian' sample has a smaller fraction of false pFoF groups, but a significantly larger fraction ($\sim 13\%$) of the pFoF groups are flagged as `serious projection'. 
Gaussian photometric redshift probability density is the simplest assumption in dealing with photometric redshift uncertainties in group finding. 
The results of Fig. \ref{gsmockplot} using `Gaussian' and `non-Gaussian' (`$z_{simulated}$') samples suggest that the asymmetric shape of galaxy's photometric redshift probability density has a role in determining group membership.

\subsubsection{Test 6: the uncertainties of photometric redshift measurement \label{mock:half}}
\indent To explore the performance of the pFoF algorithm as a function of photometric redshift measuring uncertainty, we re-construct the simulated photometric redshift sample, and then repeat Tests 1 and 2. 
We take the same steps as in $\S$ \ref{sec_mock} in generating photometric redshifts and probability densities, but reduce the dispersion between the simulated photometric redshift and mock galaxy redshift by a factor of 0.5. 
The probability densities are consequently rescaled by the same factor. 
The overall dispersion between the simulated photometric redshift and actual redshift is $\sim 0.037$ at $0 \leq z_{mock} < 0.6$ and $\sim 0.069$ at $0.6\leq z_{mock} < 0.9$. 
We apply the same criteria in selecting the sample, and refer to this sample as `$z_{half}$'. 

\begin{figure}        
	\includegraphics[width=8.2cm]{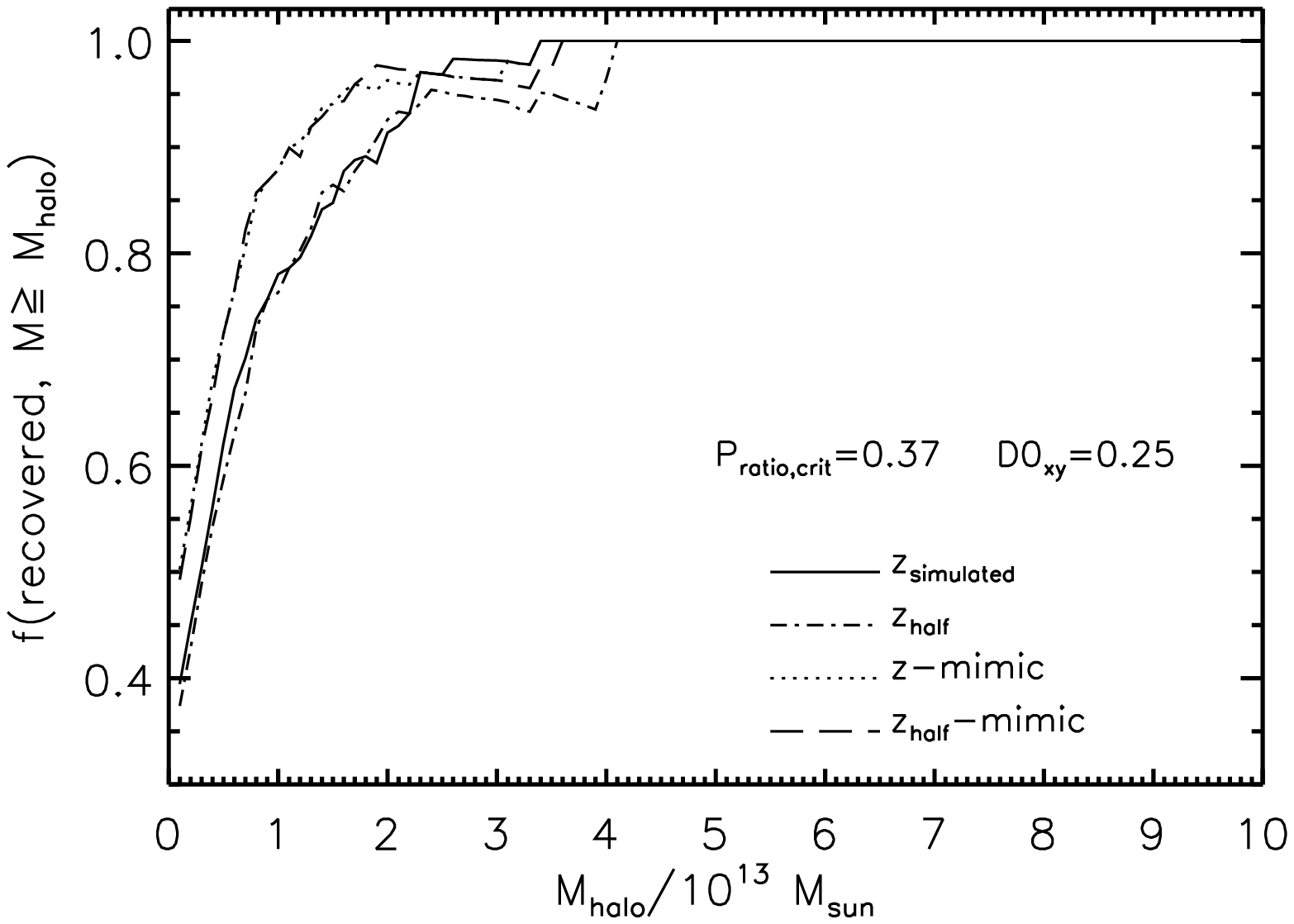}        
	\includegraphics[width=8.2cm]{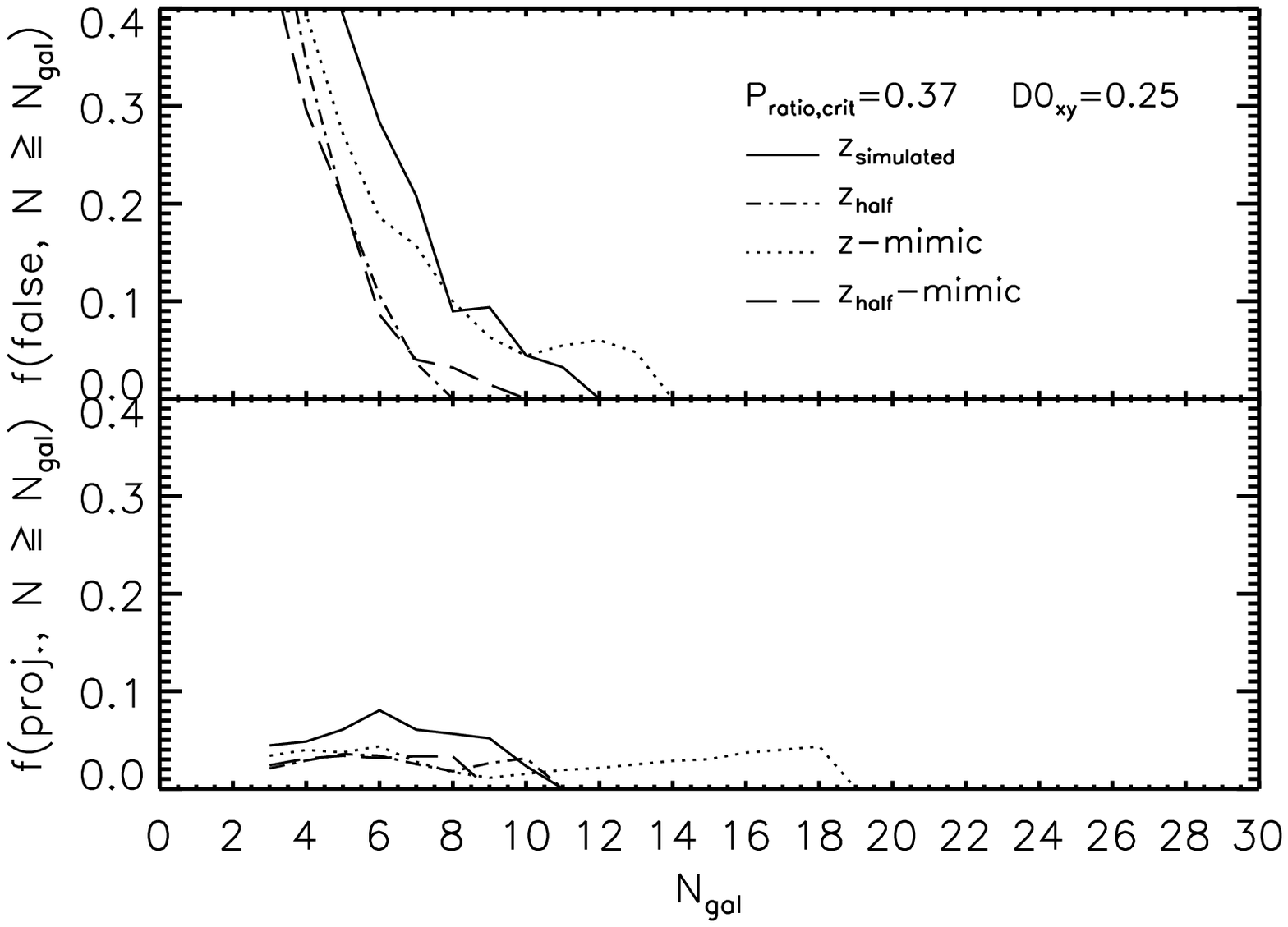}        
	\caption[Test 6 \& Test 7: accuracy of photometric redshift measurement]{\it Left: \rm The repeated Test 1 results using the `$z_{simulated}$', `$z_{half}$',`z-mimic', and `$z_{half}$-mimic' samples (see Test 6 and Test 7 in $\S$\ref{res_test}). \it Right: \rm The false detection rates (\it top\rm) and fractions of `serious projection' (\it bottom\rm) using the same four samples in the left plot. \label{FOFcomp.mock} }        
	\end{figure}

  The test results using `$z_{half}$' are presented in Fig. \ref{FOFcomp.mock} as the dash-dotted curves. 
The `$z_{half}$' sample recovers 4\% fewer mock groups of $2-4 \times 10^{13} M_{\sun}$ halo mass than the `$z_{simulated}$' (solid curve) .
However, the `$z_{half}$' sample contains a much smaller fraction of false pFoF groups -- reduced by a factor of $\sim3$ for $N_{gal}\ge6$, and equal to zero for $N_{gal}\ge8$.
Similarly, the fraction of serious projection is about 2.6\%, which is about half  the rate of the `$z_{simulated}$' sample.
This test shows that the recovery rate is not a strong function of photometric redshift uncertainty, but the false detection and serious projection rates are. 

\subsubsection{Test 7: the use of accurate redshifts \label{mock:mock}}
\indent To examine how photometric redshift accuracy affects the pFoF algorithm, we repeat Test 1 and Test 2 assigning to each galaxy its real redshift instead of the photometric one. We call this sample `z-mimic'.
The photometric redshift probability densities for galaxies in the `z-mimic' catalogs are created in the same way as the `$z_{simulated}$' sample described in $\S$ \ref{sec_mock}. 

  To test how the uncertainty in photometric redshift affects the pFoF results, we also re-construct the `z-mimic' catalogs but scale the widths of the probability densities to be half as large (i.e., by a factor of 0.5), and refer to these as `$z_{half}$-mimic' catalogs. 

  The test results using the `z-mimic' and `$z_{half}$-mimic' samples are presented in Fig.\ref{FOFcomp.mock} as the dotted and dashed curves. Both the `z-mimic' and `$z_{half}$-mimic' samples have better recovery rate ($> 80\%$ for $7\times 10^{12} M_{\sun}$) than the `$z_{simulated}$' sample ($>80\%$ for $1.2\times 10^{13} M_{\sun}$). 
The Test 2 results using `z-mimic' and `$z_{half}$-mimic' samples show that the false detection rates are $\sim 10\%$ for pFoF groups of $N_{gal} \geq 8$ and $N_{gal} \geq 5.88$, respectively. 
The `serious projection' fraction is $\sim 3\%$ on average for both samples. 
The performance of the pFoF algorithm strongly relies on the accuracy of photometric redshift measurements, as well as on the photometric redshift uncertainties of the individual galaxies (i.e., the width of the photometric redshift probability density).

\subsection{Effects of Galaxy Colors and Contamination of False Groups}
\indent  As shown in Fig. \ref{errest}, the photometric redshift uncertainties are larger for blue galaxies ($B-R_c < 1.8$) than red galaxies ($B-R_c \geq 1.8$) by a factor of $\sim$1.5 on average.
The different photometric redshift uncertainties for blue and red galaxies may result in biases in identifying galaxy groups.

To determine the significance of this effect, we test the pFoF algorithm using a `blue-improved' sample, in which we artificially make the simulated photometric redshift uncertainties for blue galaxies to be comparable to those of the red galaxies.
We find that the recovery rate is slightly better than that using the `$z_{simulated}$' sample by 2\% for groups of halo mass less than $\sim 3.4 \times 10^{13} M_{\sun}$, but the fraction of false groups is $\sim 15\%$ smaller than the `$z_{simulated}$' sample.
We note that the Test 6 results have shown a similar small improvement in the recovery rate but a significant reduction in false detection rate when all the photometric redshift uncertainties become smaller. 
Accordingly, we conclude that the larger photometric redshift uncertainties in a subset of galaxies do not affect the recovery rate, but increase the false detection rate significantly.
This is because we use photometric redshift probability densities in our group finding method, instead of using a fixed cutoff (based on some average redshift uncertainty) in photometric redshift space in determining group members. 

   One of the main issues of having different photometric redshift uncertainties between red and blue galaxies is in estimating the true fraction of the galaxy populations. 
Because galaxies in groups populate regions of relatively higher number density compared with the field, more group galaxies are expected to scatter into the field than in the reverse direction due to their photometric redshift uncertainties.
Therefore the estimated fraction of red(blue) galaxies in a group is expected to be smaller(larger) than its true value, due to the larger fraction of red galaxies in richer environments. 
To explore how significantly the true fraction is affected, we compute the fraction of red galaxies in each galaxy group. We define the red galaxies as galaxies of color redder than halfway of the $B-R_c$ color difference between E and Sc galaxies. For each recovered mock group, the true red galaxy fraction is computed simply by counting the number of red members to the total. For the matched pFoF groups in the `$z_{simulated}$' and `blue-improved' samples, we estimate the red galaxy fraction using a Bayesian inference to consider the background contamination. We find that the estimated red galaxy fraction in the `$z_{simulated}$' and `blue-improved' samples are comparable to each other. However, the values are smaller, as expected, than the true values in the recovered mock groups by $\sim 13\%$.

   Another concern in using photometric redshift groups in scientific analyses is the contamination of false groups. In a real observational sample, it is difficult to distinguish false groups from the true groups. To estimate how the contamination of false groups affects galaxy population analyses, we compute the red galaxy fraction in each pFoF group of $N_{gal} \geq 8$ for such richness cutoff. We find that the false groups have smaller red galaxy fractions compared with the true pFoF groups. The mean red galaxy fractions are $\sim 0.75$ and $\sim 0.28$ for the true and false groups, respectively. 
Therefore, when computing the averaged red galaxy fractions of all pFoF groups in a sample, the value of the estimated red galaxy fraction can be biased by $\sim 0.05$ smaller, assuming that $\sim$10\% of the groups may be false detections. 

\subsection{Examples of Recovered Groups}
\begin{figure}        
	\includegraphics[width=12.7cm]{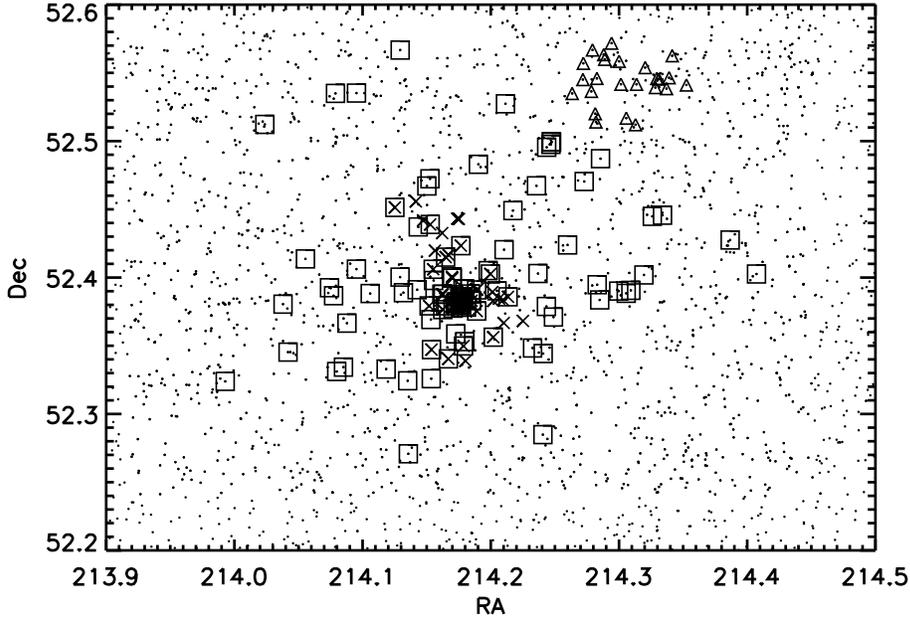}
	\caption{The sky map of a rich mock group ($z$=0.237, $M = 2.0 \times 10^{14} M_{\sun}$) and the pFoF groups in the same sky region. The solid dots are galaxies in the `$z_{simulated}$' sample. The squares mark the position of each member of the mock group in the sample. The crosses and triangles indicate the members of two pFoF groups in this region, selected with $N_{gal} \geq 10$ and $z_{pFoF} < z_{cut}$. The mock group is matched by the pFoF group plotted in crosses. Note the other pFoF group (triangles) is completely separated from the
matched one, and is identified with another mock group at $z=0.126$, demonstrating the ability of the pFoF algorithm to separate groups at different redshifts.
\label{skymap}} 
        \end{figure}

\begin{figure}
        \includegraphics[width=12.7cm]{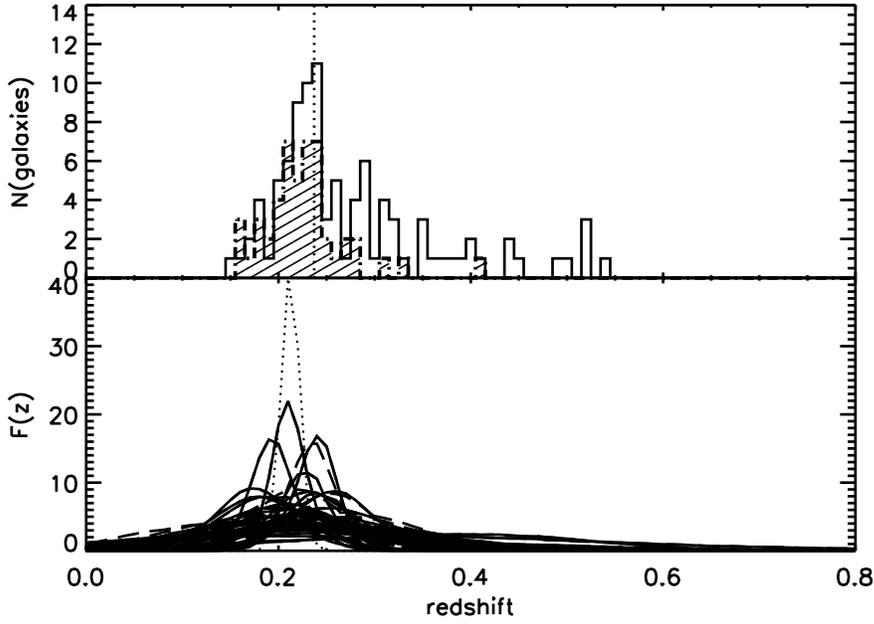}
        \caption[Redshift histogram of the rich mock group]{\it Top: \rm The histogram (0.01 bin size) of simulated photometric redshift of galaxies in the mock group (open histogram) and members of the matched pFoF group (hatched histogram). 
The vertical dotted line indicates the mock group redshift. 
\it Bottom: \rm The individual photometric redshift probability distributions of the matched pFoF group members (i.e., the pFoF group galaxies in crosses in Fig. \ref{skymap} ) are plotted as solid curves, and galaxies which belong to the pFoF group, but not in this mock group halo (i.e., galaxies in the pFoF group which are projected back/foreground galaxies) are plotted in dashed curves. The group redshift distribution of this matched pFoF group is plotted as the dotted curve, $z_{pFoF}=0.217\pm 0.009$. \label{skymap.1} }    
	\end{figure}

\begin{figure}
	\includegraphics[width=12.7cm]{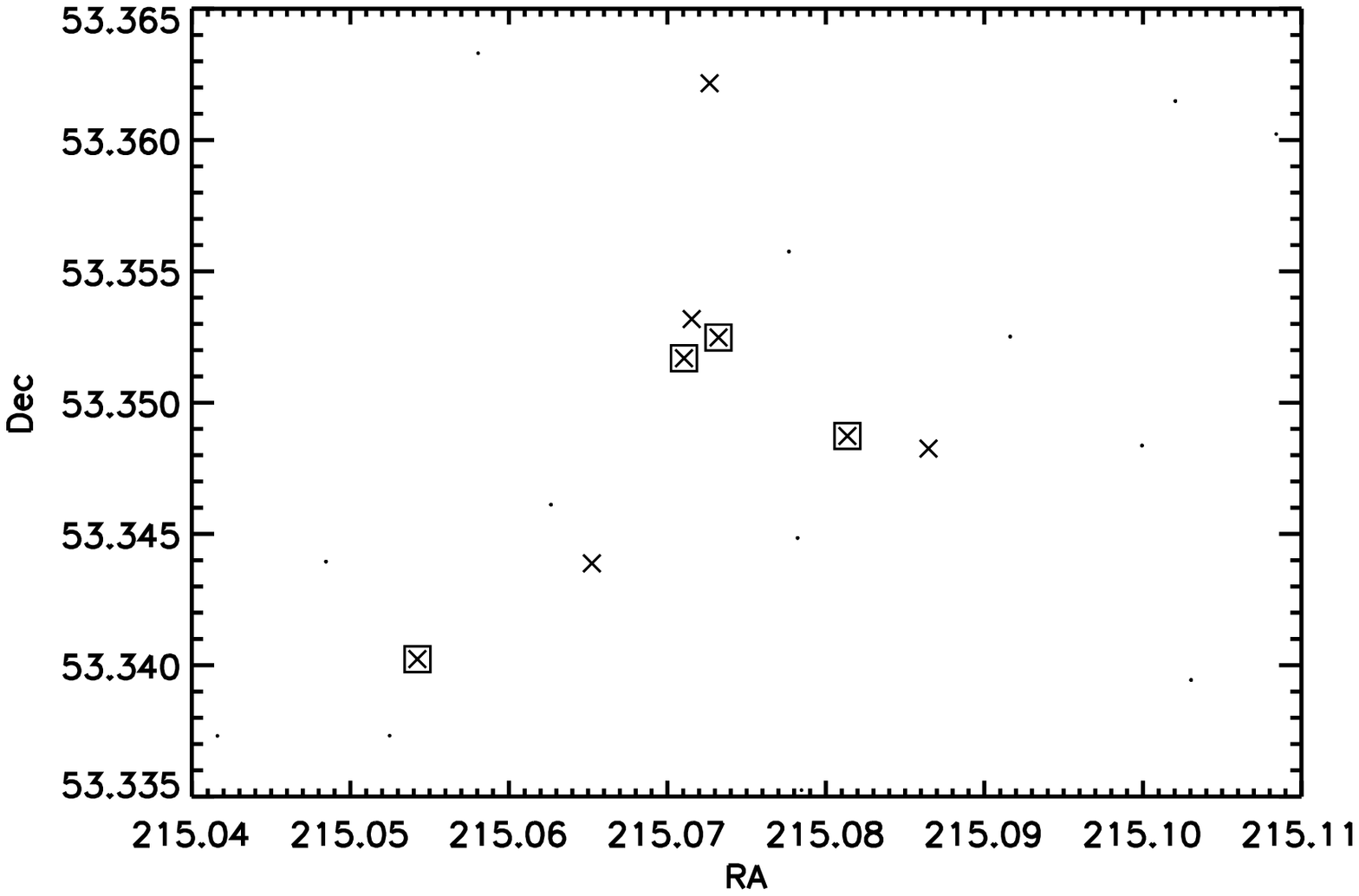} \\
	\includegraphics[width=12.7cm]{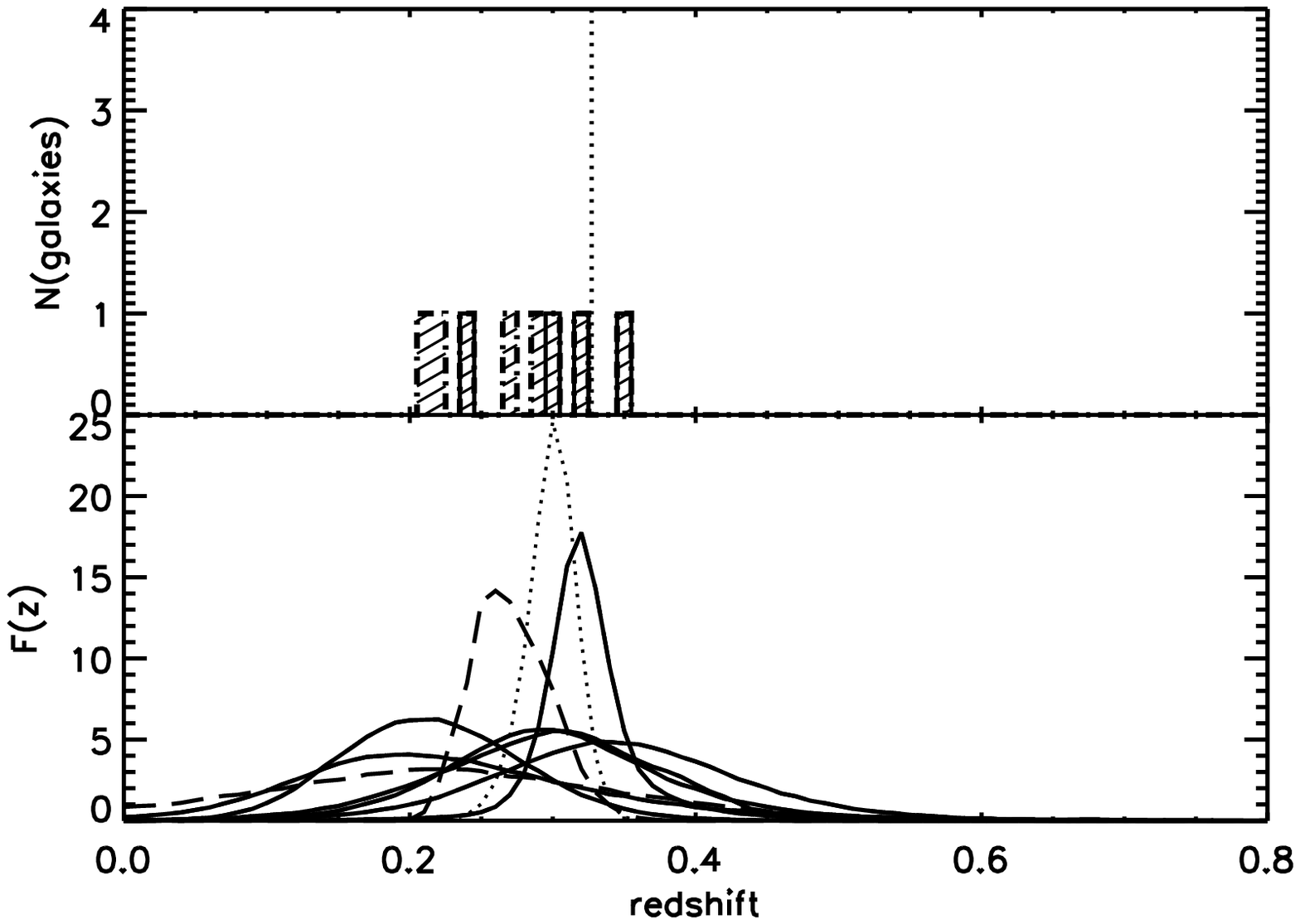}
	\caption[Example for a poorer mock group]{same as Fig. \ref{skymap} and Fig. \ref{skymap.1} but for a poorer mock group at $z$=0.327 and of $1.9 \times 10^{13} M_{\sun}$. The matched pFoF group has the richness of $N_{gal}=5.30$ at $z_{pFoF}=0.303\pm 0.018$. \label{skymap.2} }
        \end{figure}

Tests 1 to 7 allow us to conclude that our pFoF group finding algorithm is able to identify galaxy groups using photometric redshift samples, although the performance of group finding results depends on the accuracy of photometric redshift measurements. 
We summarize our test results using mock samples in Table \ref{tab:mocktest}. 
In Fig.\ref{skymap}-\ref{skymap.2}, we present two typical examples of the identified mock and pFoF groups obtained using $P_{ratio,crit}=0.37$ and $D0_{xy}=0.25$ Mpc. 
In each figure, we show the sky locations of the mock group members. 
The members of the pFoF group which matches the mock group galaxies are marked by the crosses within a square. 
The simulated photometric redshift distribution of members in the mock group and the individual photometric redshift probability densities of the matched pFoF group members are presented in Fig. \ref{skymap.1} and Fig. \ref{skymap.2}. 

Both of these figures show that the estimated pFoF group redshift probability density (the dotted curve) has a smaller width than the individual members. In fact the photometric redshift uncertainty of individual galaxies is $\sigma_{emp}\sim 0.070$, while the average estimated pFoF group redshift $z_{grp}$ uncertainty is $\Delta z_{grp} \sim 0.017$.
 This $\Delta z_{grp}$ is the width of a pFoF group redshift probability density, and depends on the number of linked galaxies. However, there is an offset between the actual and the estimated group redshifts. In our `$z_{simulate}$' sample, we find that these two sets of redshift do not follow a correlation of unity slope. 
This effect is likely related to the systematics in the photometric redshift estimation for individual galaxies. Without taking such systematic offsets into consideration, the r.m.s. of the pFoF group redshifts compared with the true ones is $\sim 0.044$. The r.m.s. is reduced to $\sim 0.020$ after correcting for such systematic offsets, and is in agreement with the estimated pFoF group redshift uncertainties $\Delta z_{grp}$. Therefore, this r.m.s. dispersion can be considered as the internal uncertainty of the redshift estimation, which is directly comparable to $\Delta z_{grp}$.

\section{Testing pFoF Algorithm on CNOC2 Groups \label{cnoc2grp}}
\subsection{The Group Samples}
   The CNOC2 group catalog was generated using a friends-of-friends algorithm with $r_p^{max}=0.25h^{-1} Mpc$ and $r_z^{max}=5h^{-1} Mpc$ as the linking parameters in the transverse and radial direction in a spectroscopic redshift sample \citep{2001ApJ...552..427C}.
A total of 192 groups in an area of 1.5 square degrees were identified at a median redshift of 0.33. 
The average number of galaxies identified in each group is $N\sim4$. 
The richness of CNOC2 groups is computed as $\eta_{CNOC2}=\sum^N_i(w_{m,i} w_{z,i})$ where $w_{m,i}$ and $w_{z,i}$ are the weights based on the magnitude and redshift selection functions \citep{2000ApJS..129..475Y}. 
As a result, the group richness is $\sim 1.74$ times greater than the identified group members, i.e., the true average group richness is $\sim 7$. 
The four CNOC2 patches coincide with the RCS1 observations (Hsieh et al. 2005), but do not have complete overlap. 
We apply the sample selection and the pFoF algorithm with $P_{ratio,crit}=0.37$ and $D0_{xy}=0.25$ Mpc to the RCS catalogs overlapping with the CNOC2 patches. 
Due to the incomplete coverages in the RCS, we have 109 of the published CNOC2 groups in our sample.
We set a redshift cut as $0.19\leq z < 0.41$, since the redshift distribution of the CNOC2 groups becomes incomplete beyond $z\sim 0.4$ \citep{2001ApJ...552..427C}.

\subsection{The Group Finding Results}
We first check the pFoF performance on CNOC2 groups and subsequently we use pFoF groups to establish the completeness of the CNOC2 group sample.

\begin{figure}
        \includegraphics[width=8.2cm]{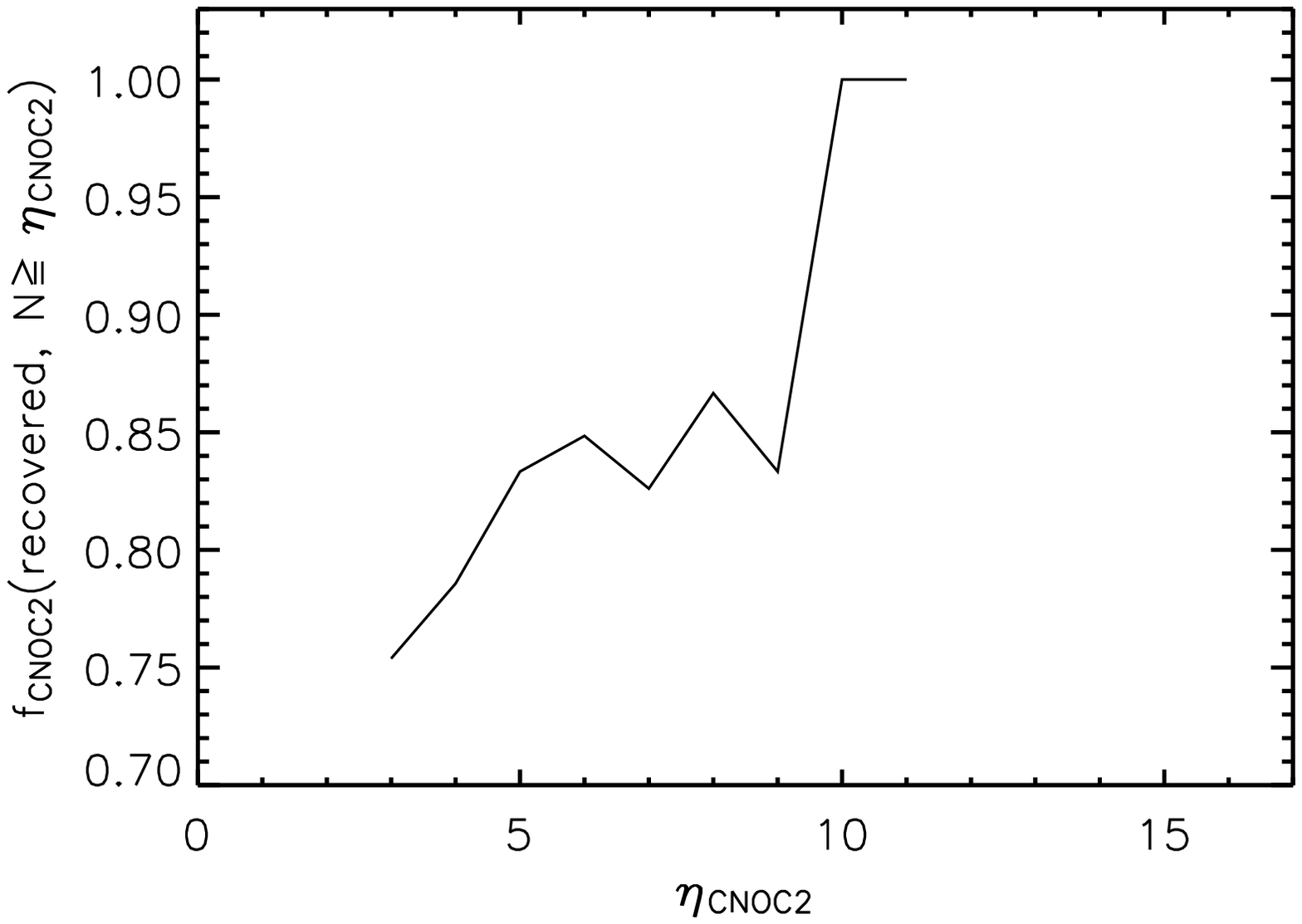} 
        \includegraphics[width=8.2cm]{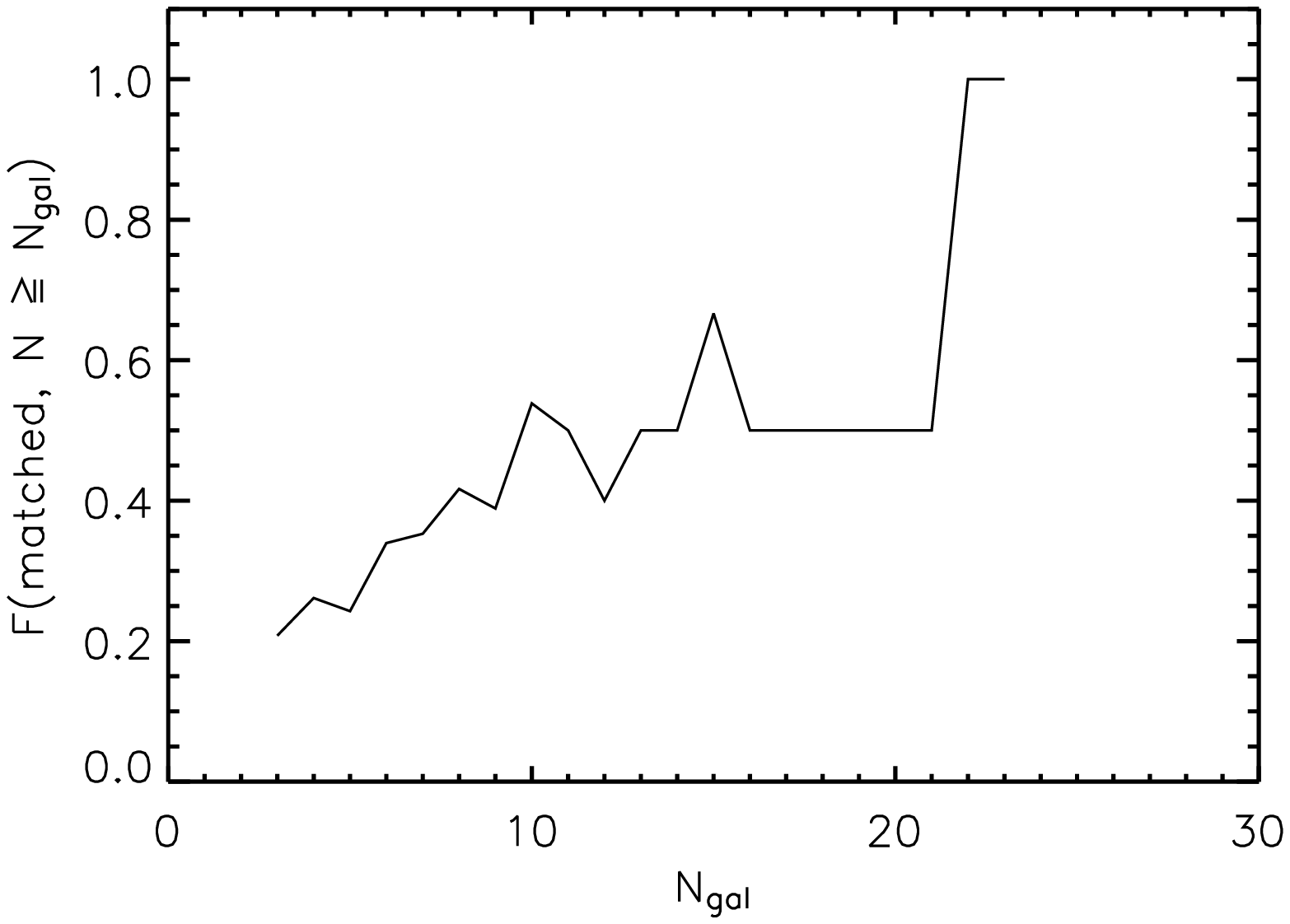}
        \caption[CNOC2 group finding results]{\it Left: \rm The recovery rate of pFoF as a function of CNOC2 group richness $\eta_{CNOC2}$. \it Right: \rm The fraction of matched reference pFoF groups to the total as a function of group richness $N_{grp}$.
	\label{CNOC2_comp} }        
	\end{figure}

\noindent (1) Test I: the fraction of recovered CNOC2 groups\\
\indent  
To establish if a pFoF group recovers a CNOC2 group, we measure the separation between the CNOC2 and pFoF group centers. 
The reference CNOC2 groups are selected with the criterion $\eta_{CNOC2}/N < 2.5$ to remove highly incomplete groups. 
With this, we have 65 reference CNOC2 groups. 
We define that a matched pFoF group must have its center within 0.25Mpc (the linking length used in Carlberg et al. 2001) to a CNOC2 group center, and satisfy $N_{gal} \geq 3$. Fig. \ref{CNOC2_comp} shows the recovery rate as a function of CNOC2 group richness $\eta_{CNOC2}$. The recovery rate is $\sim 80\%$ for the richness cutoff of $\eta_{CNOC2} \geq 5$. 

\noindent (2) Test II: the completeness of CNOC2 groups \\
\indent  To examine the completeness of CNOC2 groups, the reference pFoF groups are selected as $N_{gal} \geq 3$ and $N_{gz} \geq 5$ in the same redshift range as the CNOC2 groups. 
We have 231 pFoF groups satisfying these conditions. 
Also in this case we impose a maximum separation of 0.25Mpc between the pFoF and CNOC2 group centers.
For the purpose of estimating the sampling rate of the CNOC2 groups, we plot in Fig. \ref{CNOC2_comp} the ratio of matched reference pFoF groups to the total as a function of group richness $N_{gal}$. 
Fig. \ref{CNOC2_comp} shows that $\sim 50\%$ pFoF groups with $N_{gal} < 20$ are matched with the CNOC2 groups. 
If we take the fraction of false pFoF groups to be $\sim 10 \%$ based on the results of Test 2 in $\S$ \ref{res_test}, the result indicates that the completeness rate of the CNOC2 groups is $\sim 56\%$ for poor groups, which is similar to what was estimated (roughly 50\%) in Carlberg et al. (2001).

\section{Summary \label{sec_conclusion}}
  We have presented a new group-finding algorithm, pFoF, for identifying galaxy groups using photometric redshift catalogs. 
We have tested our pFoF algorithm on both mock catalogs and the CNOC2 groups. We summarize the most relevant results below.

 Using the sample in which the simulated photometric redshifts resemble the real data, the comparisons between the pFoF and mock groups show that our algorithm produces reasonable results: 
(1) more than $80\%$ of the mock groups with $1.2 \times 10^{13} M_{\sun}$ halo mass are recovered, 
(2) the fraction of false groups is $10\%$ for the groups of $N_{gal}\geq 7.91$, 
and (3) $\sim 5\%$ of pFoF groups are flagged as `serious projection' for which the pFoF group members are contributed by multiple mock groups. 
We find that the pFoF results strongly depend on the sample depth. 
The samples should be sufficiently deep ($\sim M^*_{R_c}+1.5$) into the luminosity function for reliable group finding results. 
The use of samples with accurate redshift measurements reveals that the false detection rate depends strongly on the photometric redshift measurement accuracy. 
Application of the pFoF algorithm to the RCS-CNOC2 patches shows good agreement for the CNOC2 groups with $0.19 \leq z < 0.41$.

The basic working principle of our pFoF algorithm is to improve the group redshift by joining new members. 
The average uncertainty in the estimated pFoF group redshift in our mock group tests is $\sim 0.017$, compared with the average uncertainty
of 0.070 for the photometric redshifts of individual galaxies.
While such group redshift uncertainty is still very large compared with groups spectroscopically identified, our results show that our pFoF algorithm reduces the photometric redshift uncertainties significantly.

With our test results, we have demonstrated that our group-finding algorithm is able to identify galaxy groups with the capability of dealing with photometric redshift uncertainties.
The purpose of this paper is to provide a method for searching galaxy groups (and clusters) in photometric redshift data sets as the first in a series of papers. We will apply this pFoF algorithm to the CNOC1 and RCS data sets. These data sets will provide us with a large sample of galaxy groups at $0.2 \leq z < 0.6$, and enable us to study environmental dependence of galaxy properties and their evolution.

\acknowledgements
We sincerely thank Darren Croton and Michael Cooper for their generous offers of mock catalogs. We also thank David Gilbank for helpful discussions. I.H.L. acknowledges financial supports from the University of Toronto Fellowship and the Helen Sawyer Hogg Fellowship. The research of H.K.C.Y. is supported by grants from the Canada Research Chair Program, the National Science Engineering Research Council (NSERC) and the University of Toronto.



        \begin{table}
        \begin{center}
        \caption{Mock Test Results \label{tab:mocktest}}
        \begin{tabular}{lllllll}
        \tableline\\
	Sample & $P_{ratio,crit}$  & $D0_{xy}$\tablenotemark{a} & $M^{k,e}_{R_c}$ & recovery rate\tablenotemark{b} & false dection rate\tablenotemark{c} & serious projection\tablenotemark{d} \\
        \tableline\\
	$z_{simulated}$ & 0.37 & 0.25 & $M^*_{R_c}+1.0$ &~ 31\% & 0\% & 8\% \\
	$z_{simulated}$ & 0.37 & 0.25 & $M^*_{R_c}+1.5$ &~ 67\% & 0\% & 5\% \\
	$z_{simulated}$ & 0.37 & 0.25 & $M^*_{R_c}+2.0$ &~ 80\% & 9\% & 5\% \\
	$z_{simulated}$ & 0.25 & 0.15 & $M^*_{R_c}+2.0$ &~ 65\% & 0\% & 3\% \\
	$z_{simulated}$ & 0.25 & 0.20 & $M^*_{R_c}+2.0$ &~ 79\% & 10\% & 3\% \\
	$z_{simulated}$ & 0.25 & 0.25 & $M^*_{R_c}+2.0$ &~ 80\% & 16\% & 8\% \\
	$z_{simulated}$ & 0.25 & 0.30 & $M^*_{R_c}+2.0$ &~ 80\% & 19\% & 8\% \\
	$z_{simulated}$ & 0.37 & 0.15 & $M^*_{R_c}+2.0$ &~ 61\% & 0\% & 5\% \\
	$z_{simulated}$ & 0.37 & 0.20 & $M^*_{R_c}+2.0$ &~ 73\% & 2\% & 2\% \\
	$z_{simulated}$ & 0.37 & 0.25 & $M^*_{R_c}+2.0$ &~ 80\% & 9\% & 5\% \\
	$z_{simulated}$ & 0.37 & 0.30 & $M^*_{R_c}+2.0$ &~ 80\% & 10\% & 7\% \\
	$z_{simulated}$ & 0.50 & 0.15 & $M^*_{R_c}+2.0$ &~ 55\% & 5\% & 1\% \\
	$z_{simulated}$ & 0.50 & 0.20 & $M^*_{R_c}+2.0$ &~ 72\% & 0\% & 3\% \\
	$z_{simulated}$ & 0.50 & 0.25 & $M^*_{R_c}+2.0$ &~ 76\% & 2\% & 2\% \\
	$z_{simulated}$ & 0.50 & 0.30 & $M^*_{R_c}+2.0$ &~ 79\% & 8\% & 7\% \\
	Gaussian        & 0.37 & 0.25 & $M^*_{R_c}+2.0$ &~ 82\% & 9\% & 13\% \\
	$z_{half}$      & 0.37 & 0.25 & $M^*_{R_c}+2.0$ &~ 80\% & 0\% & 3\% \\
	z-mimic         & 0.37 & 0.25 & $M^*_{R_c}+2.0$ &~ 90\% & 10\% & 3\% \\
	$z_{half}$-mimic& 0.37 & 0.25 & $M^*_{R_c}+2.0$ &~ 89\% & 3\% & 3\% \\
        \hline\hline\\
	\end{tabular}
	\end{center}
        \tablenotetext{a}{in Mpc}
        \tablenotetext{b}{for mock groups of $M_{halo} \geq 1.2\times 10^{13} M_{\sun}$}
        \tablenotetext{c}{for pFoF groups of $N_{gal} \geq 8$}
        \tablenotetext{d}{for true pFoF groups on average}
	\end{table}
 
\end{document}